\documentclass[sn-basic]{sn-jnl}

\usepackage{graphicx, amsmath, booktabs, bm}
\usepackage[title]{appendix}%
\usepackage{subcaption,makecell}
\usepackage{url}
\usepackage{xcolor}
\usepackage{etoolbox}

\AtBeginEnvironment{thebibliography}{\catcode`\&=12\relax}

\DeclareMathOperator*{\argmax}{arg\,max}
\hypersetup{hidelinks}

\begin{document}

\title{Why Do We Need Travel Behavior Theory in the Age of AI? Multiple Goal Pursuit as an Illustrative Theory}

\author*[1]{\fnm{Jason} \sur{Hawkins}}\email{jfhawkin@ucalgary.ca}

\author[2]{\fnm{Omid} \sur{Armantalab}}\email{oarmantalab2@huskers.unl.edu}

\affil*[1]{\orgdiv{Department of Civil Engineering}, \orgname{University of Calgary}, \orgaddress{\street{2500 University Dr NW}, \city{Calgary}, \postcode{T2N 1N4}, \state{AB}, \country{Canada}}}

\affil[2]{\orgdiv{Department of Civil and Environmental Engineering}, \orgname{University of Nebraska-Lincoln}, \orgaddress{\street{1400 R St}, \city{Lincoln}, \postcode{68588}, \state{NE}, \country{USA}}}

\abstract{Travel behavior and demand modeling seeks to understand the factors that motivate transportation decisions. At the same time, the field is increasingly adopting algorithmic and artificial intelligence (AI) tools that improve predictive accuracy, often at the cost of a grounding in hypothesis-based theory validation and behavioural explanation. In this discussion paper, I use goal pursuit theory (GPT) to illustrate why behavioral theory is a necessary complement to prediction in travel behavior research. Unlike random utility maximization (RUM) or close alternatives (e.g., random regret minimization (RRM)), GPT explicitly models how travelers (1) activate context-dependent goals (hedonic, gain, normative), (2) resolve conflicts between competing objectives, and (3) make sequential decisions across temporal scales. I demonstrate GPT's merits through three transport applications: activity scheduling (handling hierarchical goal structures), vehicle ownership (disentangling bundled mobility goals), and location choice (capturing latent goal interactions via matrix factorization). I provide actionable guidance for implementation, including: (a) hybrid choice model specifications linking goals to observable behaviors, (b) parallels to complementary behavioral theories from the transportation field, and (c) data requirements and comparative benchmarks against RUM/RRM models.}



\maketitle

\section{Context}

Travel behavior and demand forecasting research, as in many applied fields, faces new opportunities and risks with the rapid development of artificial intelligence (AI) and the growing availability of large-scale behavioral data. In one recent committee meeting, one author observed a chair predict that AI would make traditional behavioral model structures (e.g., tours and activity chains) obsolete, replaced by large AI systems trained on ``big data'' that are difficult to interpret but optimized for predictive accuracy. Such claims echo a longstanding positivist tendency in the field that privileges prediction over explanation \citep{Friedman_1953} (see also \cite{Dawes_1999, Sheppard_2001}). In ``Logics of Dislocation'', the economic geographer Trevor Barnes argues that ``[t]he real duty of the economist is not to explain our sorry reality, but to improve it.'' \cite[p. 4]{Barnes_1996}.

\begin{figure*}[h]
  \centering
    \includegraphics[scale=0.4]{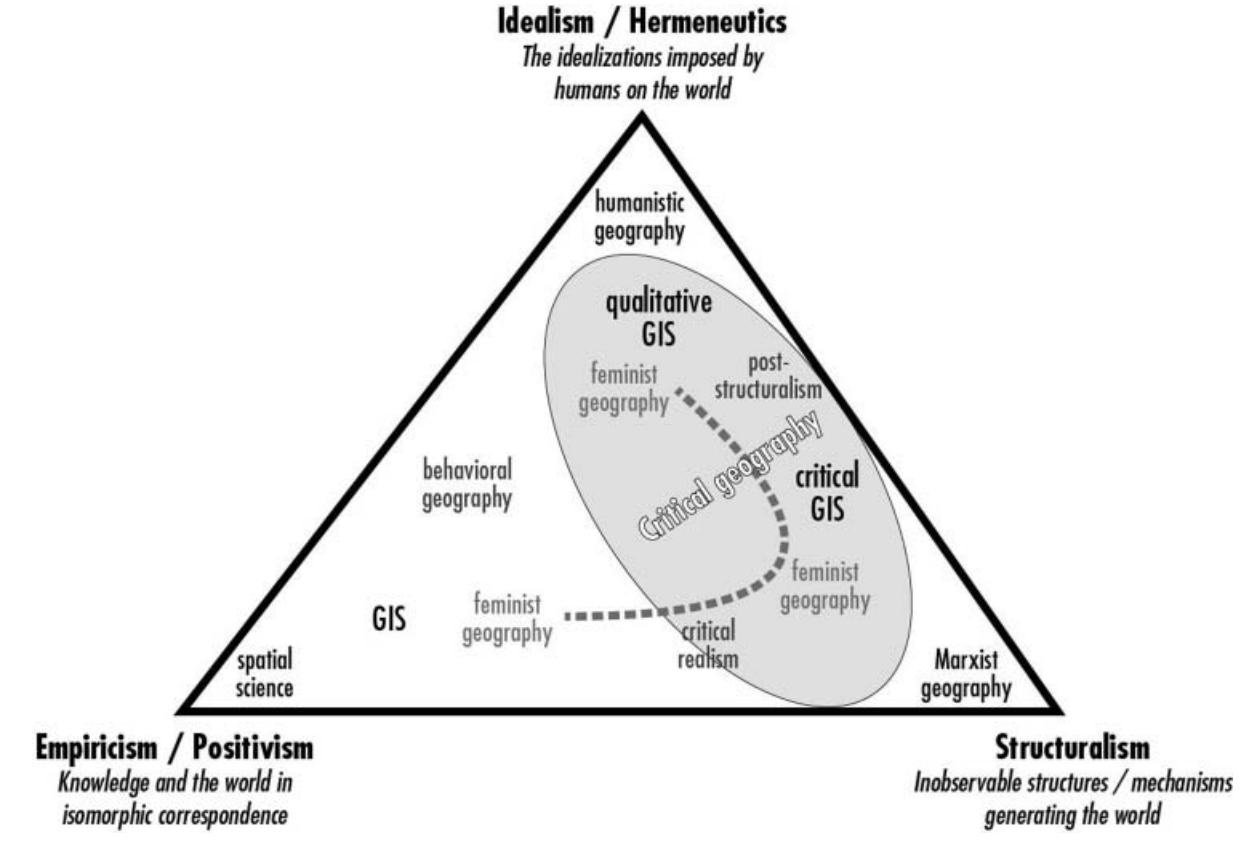}
  \caption{The epistemological triangle \citep{Sheppard_2015}}
  \label{fig:epist}
\end{figure*}

\citet{Sheppard_2015} usefully frames the discussion with an \emph{epistemological triangle}. AI tools are often positioned as methods for descriptive prediction within an existing socio-technical context (i.e., the lower-left corner of Figure \ref{fig:epist}). Prediction is valuable, but by itself it does not specify which interventions are desirable, which mechanisms drive behavior, or why changes in policy, technology, or context might produce different outcomes. Prediction is also only one tool for understanding the existing social context. It struggles to support hermeneutic interpretation and to account for structural forces embedded in prevailing economic, political, and social systems. In general, the field does relatively well at predicting event-level behavior but has a comparably poor understanding of processes and theories of behavioral change. Moving models out of the lower-left corner of Figure \ref{fig:epist} will require greater integration of theory with methods and richer data that capture systemic barriers to aligning choices with preferences.

I argue that the transportation field should not treat increasing statistical complexity as a substitute for behavioral theory. Models with comparable predictive performance can embody very different assumptions about cognition, motivation, and constraint, leading to diverging interpretations and policy sensitivities. In that sense, statistical and computational advances are best viewed as \emph{means} that can support (but not replace) the \emph{end} of behavioral understanding. This perspective has been articulated repeatedly across the behavioral and choice-modeling literatures, including the following:

\begin{quote}
... we also hope that researchers interested in decision-making and choice behavior
will recognize that statistics and econometrics, while inherently useful to our endeavors,
are not the end, but rather the means to support our continuing quest to reach the end. Far
too much effort has been expended to develop complex models for their own sake, and far
too little to develop models that parsimoniously approximate real processes. Prediction
success and good model fits do not equal understanding, and understanding is unlikely to
come from pedantically overly complex statistical models that demonstrate mathematical
and statistical ability but little understanding of theory and substance.
\hfill---Louviere, Meyer, Bunch, Carson, Dellaert, Hanemann, Hensher and Irwin (\citeyear{Louviere_Meyer_Bunch_Carson_Dellaert_Hanemann_Hensher_Irwin_1999})
\end{quote}

In what follows, I use goal pursuit theory (aptly abbreviatetd as GPT) as an illustrative example of how a richer behavioral account of motivations can complement advances in estimation and prediction. GPT has seen relatively limited empirical application in transportation to date, but it offers a structured way to theorize \emph{why} travelers make the choices they do. As stated by two of the primary developers of econometric GPT formulations:

\begin{quote}
The possibility of multiple goal pursuit in choice has been recognized in the decision making literature (Austin and Vancouver 1996), but the roles of goals as (a) metrics of attainment/evaluation, (b) constraints, (c) determinants of effort and cognitive resource allocation (Weber and Johnson 2009), and (d) mediators/moderators of interactions between [decision-makers] DM and context, have been almost universally subsumed (consciously or not) into the ubiquitous valuation/utility function. This is a heavy burden for a single construct to bear! In our opinion, this informal (even lax) treatment of goals in demand models has hampered the development of more process-rich descriptions of consumer behavior.
\hfill---\cite{Marley_Swait_2017}
\end{quote}

While I focus on a single class of theories for clarity, our intention is to motivate the continued exploration of behavioral theories, in parallel with data and model advances, to understand human decision-making. I also emphasize that this research note is intentially theoretical in its presentation in the spirit of critical social theory, leaving empirical applications for future work.

\section{From Predictive Fit to Behavioral Mechanism}

Structural\footnote{\emph{Structural} here refers to a model with explicit assumptions about the decision-maker's objectives, decision environment, and choice formation. This definition is contrasted with \emph{reduced-form} models that lack a behavioral basis and employ standard statistical model formulations. Note: a logistic model can be structural if it is founded in a behavioral theory (e.g., random utility maximization) but may also be purely statistical, depending upon the application context.} models of decision-making are commonly employed in travel behavior research. In fact, many of these models were originally formulated for travel behavior applications \citep{Domencich_McFadden_1975, Ben-Akiva_1973, Wen_Koppelman_2001}. The dominant behavioral assumption is random utility maximization (RUM), in which the decision maker selects a discrete alternative from among a finite set of available alternatives to maximize a latent utility function \citep{Potoglou_Spinney_2024}. While there are many available model structures (e.g., multinomial, nested, cross-nested, and generalized multinomial logit) and error term assumptions (e.g., logit, probit, and various mixed logit structures), adoption of alternative decision theories is less common.

Before proceeding further, we must distinguish between innovations in \emph{decision theory} and \emph{model structure}. Addressing the latter innovation first, the integration of machine learning with structural econometric choice models has received significant attention in the travel behavior research community \citep{richards2019modelling,xie2003work,zhao2010travel,hagenauer2017comparative}. These advances improve the flexibility of model formulations, thus addressing the questions of prediction accuracy and constraints imposed by structural model assumptions. However, there is no innovation on the underlying decision theory as these models typically continue to rely on the standard RUM assumption. Thus, AI may substantially expand the flexibility of the statistical mapping between inputs and choices without changing, or even making
explicit, the behavioral assumptions underlying that mapping. The challenge is therefore not simply to integrate AI with existing model structures, but to use behavioral theory to determine what processes these flexible models should represent and test.

Two prominent innovations with respect to decision theory are the adoption of random regret minimization (RRM) \citep{Chorus_Arentze_Timmermans} and the more recent adoption of mathematical psychology models (e.g., decision field theory \citep{Hancock_Hess_Choudhury_2017}, linear ballistic accumulation \citep{Hancock_Hess_Marley_Choudhury_2021}, and quantum probability \citep{Hancock_Broekaert_Hess_Choudhury_2020}). RRM is a relatively straightforward extension of RUM with respect to model estimation and interpretation. Mathematical psychology models deviate more significantly from the RUM framework and require additional work to set up the model for estimation and produce standard travel behavior metrics, such as elasticities and willingness-to-pay estimates. These psychological models hold promise for enhanced travel behavior analysis but have not reached the same level of maturity to be integrated into large-scale transport demand models. A third innovation in decision theory is rational inattention (RI), which captures the bounded rationality nature of decision-making. That is, individuals face information acquisition costs that lead to deviations from the perfect information assumption implicit in conventional RUM models. Although still following a RUM model structure, RI introduces a Bayesian prior structure to capture inattention, which leads to deviations from rational choice and perfect information \citep{Habib_2023}. However, in most instances, the transportation literature proposing alternatives to RUM theory lacks a clear grounding in travel behavior and its theory. I argue that this gap partially stems from the current epistemological focus on mathematics and empirics in the field, which contrasts with the more theoretical discourse in the behavioral geography literature of the 1960s-1980s \citep{Bunting_Guelke_1979,Cox_Golledge_1981}.

In this paper, I present an alternative set of behavioral theories, goal pursuit theory (GPT), which can be implemented using standard generalized extreme value (GEV) model structures, while still constituting a significantly different decision theory from RUM. Fundamentally, GPT posits that individuals have multiple goals they seek to achieve through the accumulation of their decisions. Empirical applications have found improved accuracy in some cases \citep{Swait_Franceschinis_Thiene_2020}. However, a discussion of GPT in the transportation behavior context should go beyond this instrumentalist framing. 

Choices are the manifestations of motivating goals \citep{Emmons_1986}, which an individual chooses to \emph{activate} within a particular choice situation. Consider a policy that seeks to decrease the environmental burden of the transportation sector. Rather than focusing solely on actions at the choice level, GPT broadens the discussion to include environmental motivations and the mechanisms for (or barriers to) activating such goals by individuals. GPT also complements time-space geography and activity-based models by providing a conceptual lens to understand why travel decisions occur in particular sequences and locations. By linking observed choices to underlying goals and motivations, GPT offers additional behavioral realism that can enrich the interpretation of activity schedules and spatial patterns captured in these models.

It is also important to situate GPT in relation to other theoretical constructs commonly employed in travel behavior research. I focus on the Theory of Planned Behavior (TPB) and the Theory of Interpersonal Behavior (TIB) as having long emphasized the roles of attitudes, norms, motivations, and habits in shaping travel behavior \citep{De_Vos_2022}. GPT does not necessarily replace these theories; rather, it provides a means to elicit and operationalize their constructs through novel survey questions and econometric model structures. Furthermore, while TPB and TIB focus on stable determinants of behavior, GPT incorporates dynamic goal activation and multi-step reasoning processes. This helps integrate behavioral theory more directly into modeling practice, complementing existing frameworks such as hybrid choice models (HCMs) and activity-based models. Tying the preceding discussion into AI applications, recent theory-guided LLM research further demonstrates that TPB constructs can be encoded within agent personas and used to structure simulated household negotiations \citep{Sun_et_al_2026}. Similar integrations could be explored between the broad GPT theory class and evolving LLM methods to ground and validate data-driven results.

This paper provides a conceptual overview of GPT and its applicability to travel behavior research. Section 2 outlines fundamental concepts in GPT. Section 3 compares GPT with RUM on theoretical and empirical grounds. I then provide three applications that implement different aspects of the theory and econometrics of goal-based choice models. The first application (Section 4) considers activity scheduling as an expression of goals operating at multiple time scales. The second application (Section 5) considers vehicle ownership as arising from the pursuit of multiple goals, the satisfaction of which are constrained by inflexibility in the choice set, with a discussion of the ability to release these constraints through the use of technology vis-\`a-vis Mobility as a Service (MaaS). The third application (Section 6) posits that a recommender system provides a structure for the duality between location choice and bid-rent models, viewed through the lens of GPT. While much of the work to operationalize goal-based models has been led by Swait and Marley \citep{Swait_Marley_2013,Dellaert2018,Blake_Dubey_Swait_Lancsar_Ghijben_2020,Swait_Argo_Li_2018,Swait_Franceschinis_Thiene_2020,Marley_Swait_2017}, I also highlight the work of other scholars, recognizing I do not provide an exhaustive discussion of all model structures and contributors to the field. The body of the paper focuses on theory and intuition, leaving mathematical details for a supplementary appendix. Our overall goal in this paper is to link GPT and its mathematical implementations to travel behavior processes. I encourage the reader to consider other travel behavior contexts and their relation to GPT, as well as other mathematical implementations of this theory beyond those discussed herein.

\section{Goal Pursuit Theory}
Understanding the goals and motivations of individual decision-makers has long been a topic of interest across many fields of study \citep{Lewin_1938,Lewin_1951}. \citet{Carver_Scheier_1998,Carver_Scheier_2011} theorize that behavior is driven by goal pursuit and the discrepancy between one's current state and their desired state. Consider the example of mode choice: an individual may be assumed to consider all the alternatives available in their region (e.g., personal vehicle, bus, light rail, biking, or walking) but face constraints imposed by external factors, such as land development density and mix, that lead to deviations from their desired choice. The GPT field defines three cognitive processes that influence decision-making, typically described as goal framings: the hedonic goal framing to feel good and self-fulfilled; the gain goal framing to optimize personal resources; and the normative goal framing to act within social and cultural norms \citep{Steg_Lindenberg_Keizer_2016, Lindenberg_Steg_2013, Légal_Meyer_2009, Bösehans_Walker_2020}. While these goals have commonalities with utility maximization, they are not fully congruent with it. In the context of travel, a hedonic goal frame might present as comfort, independence, and satisfaction with the primary transport mode. A gain goal frame assumes the traveler optimizes their money and time resources. A normative goal frame might mean a mode choice that does not maximize individual utility but aligns with social norms, such as social stigma about bus travel in the United States or a preference for train travel in Germany due to environmental concerns \citep{Beirão_SarsfieldCabral_2007,Hess_2012}. GPT is appealing because it forms a structural connection between multiple decisions through its focus on high-level goals that are satisfied through the accumulation of otherwise disparate decisions over a day, week, year, or lifetime.

GPT addresses several other limitations of standard utility theory. First, consider again the example of mode choice and its variation across trips by a given individual. One interpretation of such randomness is that it is due to unobserved variables that describe the variation in choice. \citet{Swait_Marley_2013} offer an alternative goal-based perspective that says that choice randomness arises from the pursuit of multiple goals that cannot be obtained simultaneously. As I describe below, they formulate the model as a balance between the goals of selecting the RUM best available alternative and choice diversity. A second contribution by GPT is the ability of a good, or in the case of travel and time use an activity, to satisfy multiple goals. Referred to as \emph{goal coherence} or \emph{goal congruence}, running every day can simultaneously satisfy both the goals of losing weight and of managing stress \citep{Kung_Scholer_2020,Talevich_Read_Walsh_Iyer_Chopra_2017,Sheldon_Kasser_1995,McGregor_Little_1998}. Framing the discussion in consumer theory, one can consider the running activity to be substitutable for other activities in its instrumental benefits, similar to how one might consider it substitutable with respect to its intrinsic characteristics. 

Another critical component of GPT is the distinction one can draw between alternative goal structures. \citet{Kung_Scholer_2020} define three goal structures: hierarchical, network, and sequential. A hierarchical structure classifies goals based on their level of abstraction. Higher-order goals (e.g., values and identities) dictate the weighting of lower-order goals accomplished through daily tasks. Conversely, lower-order goals serve as the means to satisfying higher-order goals. A network structure classifies goals based on the strength of their association \citep{Sattath_Tversky_1977}. \citet{Förster_Liberman_Friedman_2007} show that the activation of goals exhibits a network structure, making the activation of goals more (or less) likely conditional upon the activation of associated goals. As detailed below, \citet{Swait_Marley_2013} provide an econometric model framework in which goals are \emph{activated} based on the feasibility of achieving each goal. Furthermore, Marley and Swait implicitly capture hierarchical goals in their econometric work by drawing a distinction between \emph{abstract goals} and \emph{functional goals}. Finally, a sequential structure arranges goals based on their temporal sequencing---e.g., accomplish goal 1, then accomplish goal 2, and finish.

Construal level theory provides a psychological explanation for this hierarchical relationship. Psychologically distant decisions---whether temporally, spatially, socially, or hypothetically distant---tend to be represented through abstract, high-level construals emphasizing superordinate ends, such as goals, values, and identities. Psychologically proximal decisions are more likely to be represented through concrete, low-level construals emphasizing subordinate means and immediate feasibility considerations \citep{Trope_Liberman_2010}. In travel behavior, long-term decisions such as
home location or vehicle ownership may therefore be construed in terms of lifestyle, identity, or accessibility goals, whereas an imminent mode, departure-time, or route choice may foreground concrete considerations such as travel time, distance, and cost.

\begin{figure*}[h]
  \centering
    \includegraphics[scale=0.4]{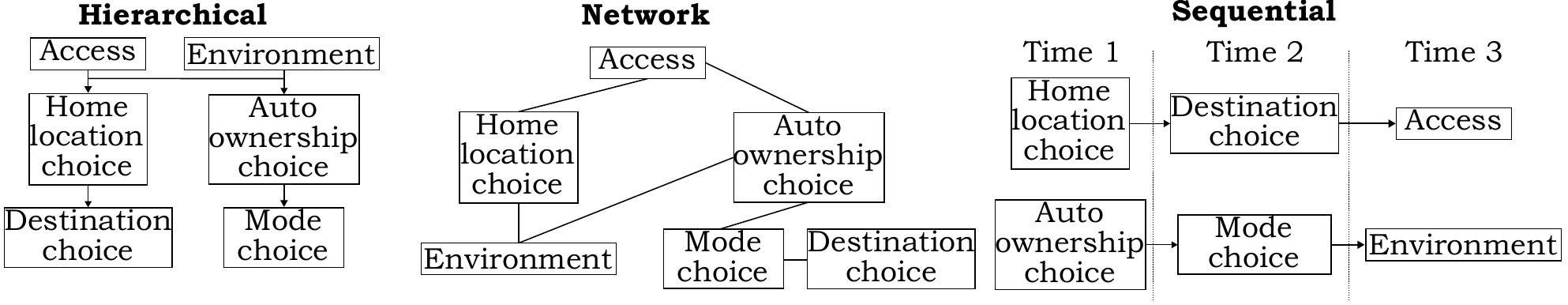}
  \caption{Principles and examples of three goal structures (adapted from \cite{Kung_Scholer_2020})} %
  \label{fig:mgp_structure}
\end{figure*}

I illustrate the three goal structures in Figure \ref{fig:mgp_structure} for transportation choices. Higher-order goals might include (1) the desire to minimize the environmental impacts of one's consumption and (2) maximize access to amenities. These goals would be realized through the choice of home location and auto ownership, which would then influence the daily pursuit of goals associated with destination and mode choices. Taking a network goal perspective, auto ownership has a strong association with mode choice, while both auto ownership and home location choices have strong associations with environmental goals. For example, an individual might choose to own a single vehicle to minimize their environmental burden but select the driving mode over lower impact modes for a given trip based on their destination choice. While the set of choices in this example do not perfectly align with a sequential goal structure, I maintain them for consistency across the figure. Home location and auto ownership are long-term decisions, which will affect daily destination and mode choice. The first application to activity scheduling provides a more complete illustration of the sequential goal structure. It is worth highlighting that while nested logit models or activity-based models already capture linked or hierarchical decisions, GPT differs conceptually by framing these decisions explicitly around goal activation and trade off between multiple objectives, allowing for a more behaviorally grounded interpretation of decision-making processes. 

GPT comprises a diverse set of theories that can structure travel behavior hypotheses. From an econometric implementation perspective, one can consider a goal-based choice model to comprise two stages \citep{Dellaert2018}. In the first stage, \emph{goal evaluation}, alternatives are evaluated against the goals set by the individual---e.g., Is a food healthy? Is it tasty? Is it cheap? This first stage can be conceptualized as a choice set generation process. In the second stage, \emph{goal choice strategy}, alternatives are measured through a balancing of these goals and an alternative chosen---e.g., Should I focus exclusively on the healthiness of a food or a balance of healthiness and cost? Existing econometric specifications are provided in appendices.

\section{Comparison of GPT and RUM}
\begin{figure*}[h]
  \centering
    \includegraphics[scale=0.4]{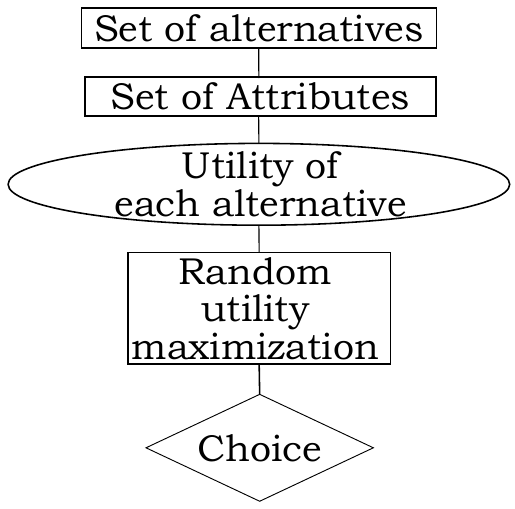}
  \caption{Standard discrete choice model structure} %
  \label{fig:traditional_dc}
\end{figure*}

GPT can be built up from Lancastrian utility theory, which underlies much travel behavior research. \citet{Lancaster_1966} notes that consumer theory at the time of his writing did not have an appropriate mechanism for considering the consumption of diverse goods---e.g., it could account for the substitution between butter and margarine but not shoes and ships. Lancaster proposed a solution based on the preference ordering of collections of characteristics rather than collections of goods - i.e., indirect utility. Choice is then defined as a function of the sum of utility partworths and the comparison of summed utilities \citep{Green_Srinivasan_1978,Guadagni_Little_1983,Troutman_Shanteau_1976}. Figure \ref{fig:traditional_dc} provides a standard discrete choice model structure (\`a la Lancastrian utility theory) from which I will build up the goal-based models. The model structure is highly abstracted to illustrate only the core elements of alternatives, described by their attributes, which enter random utility functions that are maximized.

\subsection{Empirical Comparison}
The limited empirical applications of GPT correspondingly limit our ability to compare model prediction accuracy to standard RUM-based models. \citet{Hur_Allenby_2022} find that a GPT-based model of random regret minimization (RRM) constrained by RUM (to be explicated further below) outperforms pure RRM and RUM models based on log-marginal density (LMD) and Watanabe Akaike Information Criteria (WAIC) criteria. \citet{Swait_Franceschinis_Thiene_2020} find that a hybrid choice model (HCM) specification based on GPT outperforms mixed logit (MXL) and latent class (LC) models based on AIC and BIC criteria. \citet{Swait_Argo_Li_2018} also find that their adaptive GPT-based model outperforms an LC specification based on BIC comparison across latent goal (for GPT) and class (for LC) counts.

\subsection{Theoretical Comparison}
Conventional model comparison in travel behavior research is conducted via metrics of prediction accuracy. This convention is grounded in the positivist epistemology, as exemplified by Friedman's ``The Methodology of Positive Economics'' \citep{Friedman_1953}. Let us then return to \citet{Lancaster_1966} , who provided the example of a dinner party to illustrate indirect utility. Treated as a single good, a meal possesses nutritional and aesthetic characteristics, which will differ across alternatives in their relative proportions (see our goal-based discussion in the introduction). A dinner party is then the combination of a meal and a social setting, possessing a combination of nutritional, aesthetic, and intellectual characteristics. These characteristics may differ in the combination of these two goods relative to their independent consumption. 

GPT, as detailed by \citet{vanOsselaer_Janiszewski_2012}, differs from Lancastrian utility in its assumption of consumer valuation. While Lancastrian utility ascribes an \emph{intrinsic} value to goods, GPT assumes that consumers value the benefits provided by goods and not their attributes - i.e., goal-based utility is based on \emph{instrumental value}. The instrumental value perspective of GPT parallels the \emph{derived direct demand} nature of travel. That is, in most cases, an individual values the relocation afforded by transport in a vehicle, not the travel activity, per se. If we had access to teleportation, there would be minimal loss of intrinsic value associated with the transportation system, except in cases of transport through aesthetically pleasing environments. Returning to the three goal framings given above, GPT posits that benefit can be measured through positive affect (emotion) associated with behavioral outcomes. Goal framings can be extended to include achievement, social status, emotional (e.g., happiness, excitement, relaxation), and physiological (e.g., hydration and alertness) goals \citep{Cannon_1932,Maslow_1943}. Choice is the selection of one or more means (i.e., goods or services) evaluated based on goals (i.e., benefits) to reach an outcome representing the accumulation of means, the summation of goal-weighted instrumentality \citep{vanOsselaer_Janiszewski_2012}. Choice in a given situation is then tied to goals and motivations arising from past outcomes - success or failure \citep{Emmons_1986}.

\section{Application 1: Activity Scheduling as a Hierarchical and Sequential Goal Problem}
A fundamental challenge in the transportation modeling field is how to represent the dynamics between long-term decisions about home location, employment, family formation, and auto ownership on the one hand and short-term decisions about daily mode, destination, and route choice on the other \citep{Salvini_Miller_2005,Adnan_2016,lopez2015time,Hawkins_NurulHabib_2024,Eliasson_Mattsson_2000}. From the GPT perspective, these dynamics represent a hierarchical goal structure. Activity scheduling---the planning and organization of daily tasks and events---is the daily manifestation of this decision process and itself constitutes sequential goal pursuit. Construal level theory further suggests that an activity may initially be planned in terms of abstract goals but become increasingly governed by concrete constraints as its execution approaches \citep{Trope_Liberman_2010}.

This temporal distinction also suggests the need for judicious application of cognitive-psychological theories in travel behaviour research. Theories developed to explain responses to bounded and immediately presented choice stimuli may be most informative for proximal decisions, such as route, departure-time, or human factors applications examined via driving simulator experiments. Strategic choices concerning residential location, employment, family formation, or vehicle ownership unfold over longer periods and are shaped by evolving goals, institutional constraints, social relations, and prior decisions. Treating such choices as isolated stimulus--response events risks misinterpreting the mechanisms driving these processes. Cognitive mechanisms may still contribute to their explanation, but should be embedded within theoretical frameworks that represent hierarchical goals and behavioural change across temporal scales. This application therefore explores both hierarchical and sequential goal structures.

Considering first the sequential goal structure, activity scheduling is the process of selecting which activities to conduct on a given day and for what duration. Let us further consider activity scheduling over the course of a week. \emph{Goal conflict} is a germane concept here - i.e., how an individual responds when faced with two goals to accomplish on a given day. \citet{Kung_Scholer_2020} identify three strategies in response to goal conflict: choosing, prioritizing, and multitasking. An individual can choose to accomplish one goal based on maximal utility or moral principles. Other goals are ignored in a process termed \emph{goal shielding} \citep{ShahFriedmanKruglanski2002}. Alternatively, they could prioritize goals and complete them in sequence (potentially on different days). For example, perhaps a person needs to purchase both groceries and supplies for a home repair but has time to complete only one trip. They have a dinner party that evening requiring groceries, whereas home repairs can wait until the following day, so they decide to delay their trip to the hardware store for another day. While single-day travel diaries are the norm, there is a growing literature on multi-day travel diary data \citep{Winkler_Meister_Axhausen_2024,Islam_Habib_2012}. As these datasets become more common, it becomes feasible to model inter-daily activity scheduling and its associated trade-offs. 

In contrast to sequential goal pursuit, multitasking involves the simultaneous pursuit of multiple goals. Two forms of multitasking are defined in the literature \citep{Wickens_McCarley_2019}. \emph{Concurrent multitasking} comprises the pursuit of multiple goals through a single task. For example, an individual may be driving and talking on their phone\footnote{This practice is not endorsed by the authors.}. Alternatively, an individual may perform \emph{task-switching} (or \emph{interleaving}) in which they alternate between two tasks. An example of task-switching relevant to activity scheduling and travel behavior is the work-from-home choice. An individual may choose to work from home to satisfy work and familial goals, though possibly at the expense of a reduced capacity to satisfy either goal \citep{Rubinstein_Meyer_Evans_2001}.

Turning now to the hierarchical goal structure and its applicability to activity scheduling, I will outline my first goal-based econometric model. However, let us first motivate the model development with a brief discussion of relevant travel behavior theory. \citet{Pred_1981} defines a \emph{project} as ``the entire set of tasks necessary to the completion of any goal-oriented behavior" (p.234). He situates \emph{projects} within time-space geography as spawning activities constrained by the available time-space prism. \citet{Axhausen_1998} reintroduced the concept of \emph{projects} to the travel behavior literature as long-term goals forming ``a complex system of constraints" (p. 7) on the activities one engages in over a day. Note that \citet{Palys_Little_1983} also introduced the \emph{project} term to the psychology literature as a means of understanding perceived life satisfaction and the sociologist \cite{Gershuny_2000} used the term in his discussion of time use. \citet{Miller_Roorda_2003} operationalize the project concept into an activity-based model (TASHA) as a framework for activity scheduling. \citet{Pred_1981} further distinguishes between household and institutional/organizational goals---e.g., if a project is set by an organization, then an individual will be unable to delegate their role unless they are the boss. An intriguing avenue for further exploration is the fixity of organizational goals with increased work-from-home patterns, whereby the home and workplace are coincident and time-space constraints are shifted. I term this relationship as the \emph{life path-daily path dialetic;} your life path is influenced by your daily path (i.e., activity schedule) and intersections with other life paths that expose you to new choice alternatives and affect your life path choice.

In other work, \citet{Schlich_Axhausen_2003} discuss the concept of habitual travel behavior, which can be related to goal activation. Habitual behavior is defined in the psychology literature as the automatic repetition of past actions due to learned associations between actions and outcomes \citep{Perez_Dickinson_2020,Dickinson_1985}. Habitual travel is therefore disassociated from its initial goal and may become suboptimal if the environment (i.e., traffic conditions) changes. For example, a person may have chosen their route to work to minimize travel time. In the long-term, traffic may increase on that route, leading to congestion and a different route having the shortest travel time. \citet{Kruglanski_Szumowska_2020} argue that such habitual behavior may still be goal-driven if we consider a change in the pursued goals - i.e., in the prior example, the travel time minimization goal is replaced by a route search time minimization goal. Along a parallel vein, \citet{vanOsselaer_Janiszewski_2012} suggest that goal activation may comprise chronic and temporary components. Chronic activation measures the stable long-term goal weighting that arises from personal background and culture. Temporary goal activation is a consequence of goal priming from the daily scheduling process, such as previous activity goal (non)achievement and the time between activity engagement decisions.

\begin{figure*}[h]
  \centering 
    \includegraphics[scale=0.4]{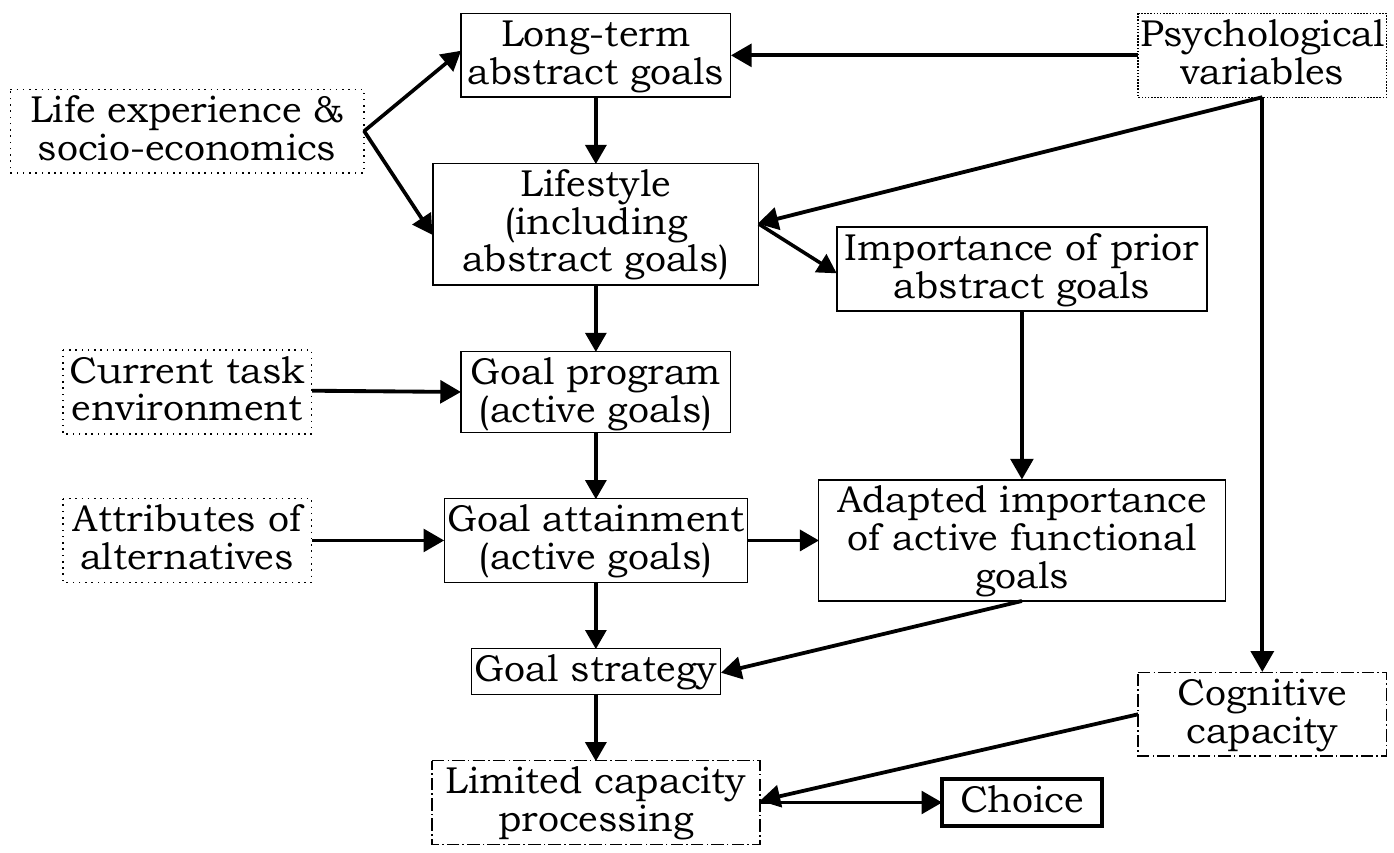}
  \caption{A path diagram of a goal-based latent variable and choice model (adapted from \cite{Marley_Swait_2017})} 
  \label{fig:latent_structure}
\end{figure*}

The full goal-based hybrid choice model (HCM) structure is illustrated in Figure \ref{fig:latent_structure} as a conceptual model, components of which have been implemented in various forms \citep{Marley_Swait_2017, Swait_Franceschinis_Thiene_2020}. Capacity limits can be implemented in the model through scale parameter heterogeneity, which is interpreted as stemming from individual characteristics, including cognitive capacity constraints. Note that cognitive capacity effects in the goal-based HCM model parallel the motivation for rational inattention \citep{Habib_2023} and there may be opportunities to incorporate the two concepts within an econometric model structure.

\begin{figure*}[h]
  \centering
    \includegraphics[scale=0.4]{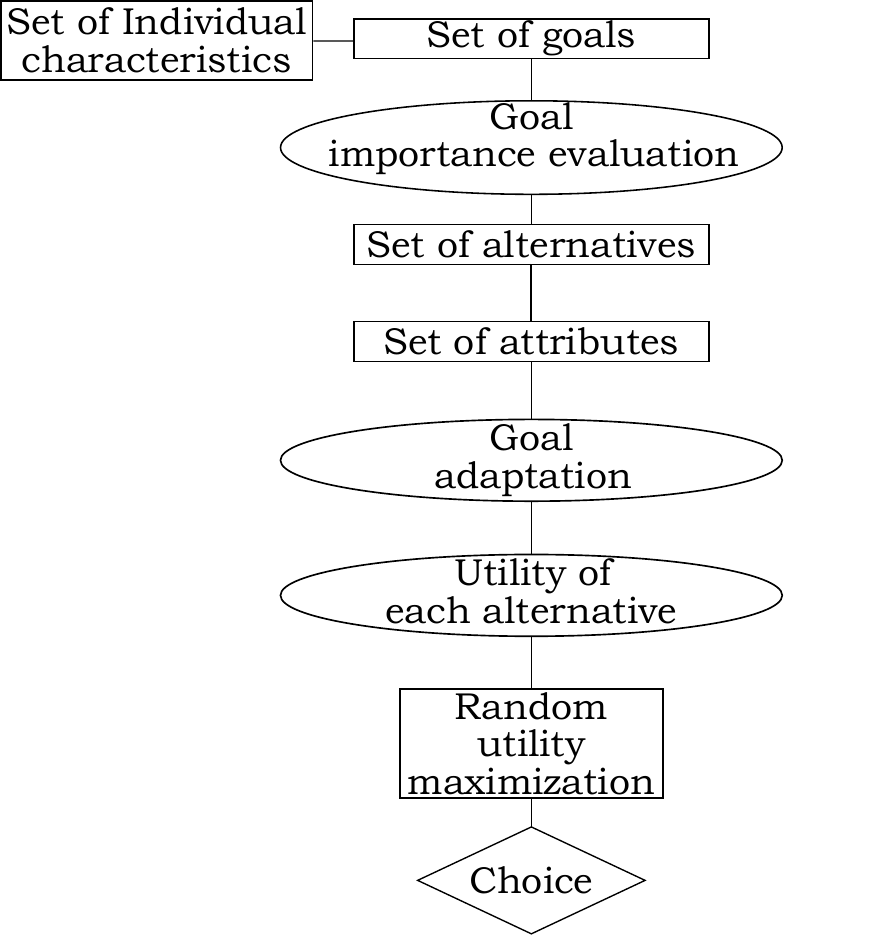}
  \caption{Goal-based discrete choice model structure} %
  \label{fig:goal_attainment_dc}
\end{figure*}

Figure \ref{fig:goal_attainment_dc} provides a simplified depiction of the goal-based extension to the standard discrete choice model. Goal evaluation has been incorporated into the model in two ways. The first model formulation takes a hybrid choice model (HCM) structure, with goals replacing the attitudes in the standard model specification. GPT helps to formalize goal structures and mental models within the HCM modeling framework. The model can be separated into goal space and attribute space components \citep{Swait_Franceschinis_Thiene_2020}. In the goal space, individual attributes describe the importance of each goal through a goal activation function. Goal importance is defined as latent variables through the use of a Likert scale and ordered process model. These goals then enter the attribute space choice model as latent variables in the utility function. \citet{Kim_Bailey_Hardt_Allenby_2017} develop a latent benefit-based conjoint analysis with a similar structure, if we interpret their \emph{latent benefits} as goal constructs. However, unlike hybrid choice or latent-variable models, which incorporate psychological constructs as predictors of utility, GPT frames decision-making around the activation and prioritization of multiple goals, providing a behavioral mechanism that extends beyond the statistical role of latent variables.

The second model formulation is a 2-stage choice model, wherein long-term abstract goals are dealt with in a choice set formation stage, following which short-term functional goals are then pursued at the choice stage within the subset of considered alternatives \citep{Marley_Swait_2017}. The choice set model defines a goal activation function and the probability that a particular subset of alternatives is considered under the chosen goal strategy. The choice model then captures a process of goal adaptation based on the context of the choice situation. Goal attainment and mixing of goal attainments can be captured using archetypal analysis, which is based on the idea that multidimensional features of the decision-maker can be interpreted as a mixture of extremal points (pure archetypes). \citet{Swait_Argo_Li_2018} use two archetypes, termed \emph{no-adaptation} and \emph{adaptation} types. We can also consider goal thresholds as a screening mechanism in the choice set stage. With an available set of alternatives, the attainability of a goal by a given alternative is a function of its ability to attain a minimum threshold level. A standard threshold function is based on cumulative prospect theory, whereby deviations below the threshold are weighted more heavily than deviations above it \citep{Tversky_Kahneman_1992}. In another study by \citet{fu2017two}, a two-stage network-based modeling approach is used to study customers' consideration and choice behaviors separately but simultaneously. In the first stage, customer preferences are modeled to form a consideration set of multiple alternatives, while in the second stage, the customer's choice preference is modeled once the consideration set has been formed. This approach ties the model framework to the third (network) goal structure. From the network perspective, goals are arranged by degree and strength of associations \citep{Sattath_Tversky_1977} and activation of a given goal can make it more or less likely that a connected goal is also activated \citep{Förster_Liberman_Friedman_2007}.

Latent variable models are commonly applied to understanding vehicle ownership, and the effect of decision-maker attitudes and perceptions on the choice \citep{yan2022empirical}. \citet{mohammadi2021investigation} use an integrated choice and latent variable (ICLV) model to understand the influence of safety attitudes on vehicle type choice. \citet{nazari2024electric} use an ICLV model to investigate the adoption behavior of electric vehicles and vehicle transaction decisions. They incorporate latent variables related to attentiveness to vehicle attributes, social influence, environmental consciousness, and technology savviness into the model. \citet{sharma2020modeling} use an ICLV model to examine how individuals' social networks and values influence their decisions regarding connected and automated vehicles (CAVs). However, none of these studies make an explicit reference to decision-maker goals. Two-stage choice models have also been applied to the vehicle ownership problem. \citet{xu2024agent} split the decision process into a choice set generation stage that narrows the considered alternatives from the entire auto market and a second stage vehicle choice model. \citet{bi2021modeling} investigate how consumer preferences change from consideration in the choice set stage to the final choice. Again, these studies do not make explicit reference to goals. The mode choice literature includes more explicit reference to goal pursuit. \citet{Geng_Long_Chen_2016} consider the influence of environmental information on mode choice via a controlled trial. They define six travel goals and perform clustering analysis. \citet{Paulssen_Temme_Vij_Walker_2014} estimate an ICLV model of mode choice, which is motivated by the effect of values (i.e., goals) on the choice . They consider power, hedonism, and security values, which align with the GPT inclusion of hedonism as a goal framing \citep{Steg_Lindenberg_Keizer_2016}. The above-described HCM and two-stage models are readily applicable to vehicle ownership choice to make a more explicit connection with GPT. In the next section, I consider the vehicle ownership decision using a different goal-based modeling framework for the case of electric vehicle (EV) adoption.

\section{Application 2: Disentangling Vehicle Choice}
Among the most cited reasons for reluctance to buy an EV is range anxiety \citep{Pevec_Babic_Carvalho_Ghiassi-Farrokhfal_Ketter_Podobnik_2019}. According to \citet{USDepartmentofEnergy_2024}, the median EV range for model year 2024 was 283 miles compared to 403 miles for internal combustion engine vehicles (ICEVs). While the EV sector could pursue driving range parity to compete with ICEVs, greater driving range typically translates to heavier and more expensive vehicles. The powertrain comprises 51\% of the purchase price for an EV compared to 18\% for an ICEV \citep{HenriqueRuffo_2020}. Furthermore, additional battery weight results in more severe traffic crashes and greater pavement damage \citep{Shaffer_Auffhammer_Samaras_2021}. The simplest way to reduce EV price and weight is to shrink battery size, thus also reducing driving range (holding battery technology constant). Developments in battery and charging technology are leading to significant reductions in charging time, turning an hour stop into a 5-10 minute one \citep{Maloney_2022, Zhang_2025}. These changes have the potential to shift the EV charging paradigm toward one more similar to ICEV refueling. I argue below that the underlying travel behavior problem is that vehicle choice typically ignores the entangled issue of vehicle and travel choices. Goal-based models offer a way to address this constraint on travel behavior analysis.

An individual faces a set of vehicle alternatives with various attributes (e.g., price, mileage, powertrain, number of seats). The standard choice model treatment includes these vehicle attributes in a single utility function. However, vehicle ownership comprises a diversity of goal pursuits that the individual must balance including travel, environmental, and vehicle attribute-related goals. Satisfying the travel goal might require a vehicle with a long driving range. An environmental goal to reduce greenhouse gas (GHG) emissions may be constrained by household income and the price of an EV relative to an ICEV (though EV prices continue to drop \citep{Randall_2024}). The vehicle attribute goal might relate to having sufficient space for all household members in the vehicle. Let us make the modeling problem more explicit by considering the personal vehicle as only one part of a larger \emph{mobility bundle}. We can then disentangle vehicle attribute choices from the travel goals, such as occasional long-distance trips, which might be met by a combination of travel modes. In operational terms, we can think of this disentangling of the personal vehicle into a bundle of services that are centrally coordinated through a digital platform - i.e., Mobility as a Service (MaaS) \citep{Hensher_Mulley_Nelson_2023}.

Vehicle ownership also illustrates how goals may be socially mediated. Drawing on mimetic theory, \citet{Burgis_2021} argues that individuals often learn what to desire by observing social ``models'' whose choices signal desirable identities and lifestyles. A vehicle may therefore be valued not only for its functional attributes, but because ownership is associated with the status, independence, environmental identity, or technological sophistication represented by an admired reference group. Mimetic desire operates upstream of choice: exposure to peers, neighbours, advertising, or online communities may affect which vehicle-related goals are activated, how strongly they are weighted, and which alternatives are perceived as capable of attaining them. This perspective distinguishes socially formed goals from direct conformity to a norm; consumers may imitate an aspirational model even when the modeled choice is uncommon within their immediate community.

\begin{figure}[h]
    \centering
    \begin{subfigure}{0.48\textwidth}
        \centering
        \includegraphics[scale=0.4]{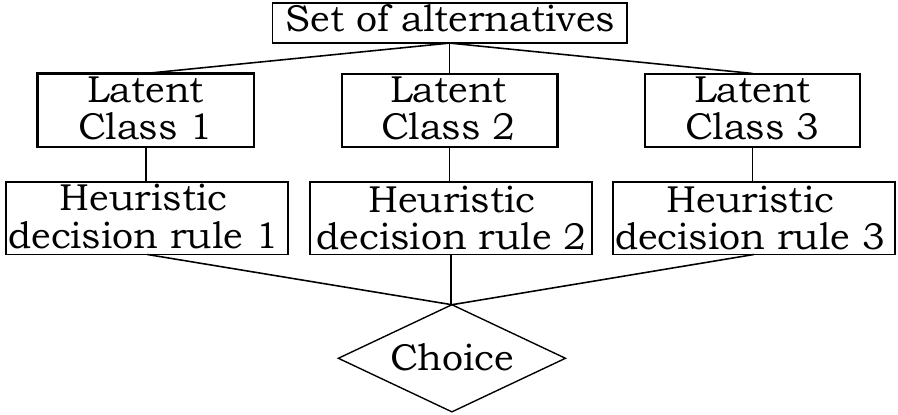}
        \caption{Latent class model structure}
        \label{fig:latent_class_dc}
    \end{subfigure}
    \hfill
    \begin{subfigure}{0.48\textwidth}
        \centering
        \includegraphics[scale=0.4]{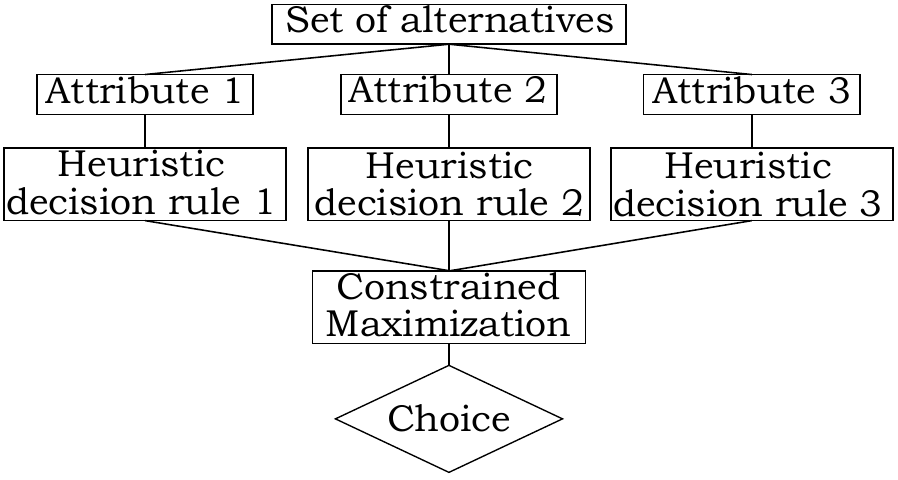}
        \caption{Constraint-based model structure}
        \label{fig:constraint_dc}
    \end{subfigure}
    \caption{Multiple heuristic discrete choice models}
    \label{fig:multiple_heuristic_dc}
\end{figure}

In this application, I will define goals by alternative decision heuristics---e.g., random utility maximization, random regret minimization, and elimination by aspects. There are two approaches to multiple decision heuristics that must be distinguished between before proceeding further. In the first approach, latent class models (LCMs) are used to capture heterogeneity in the decision rules applied by individuals (see \cite{Kim_Mokhtarian_2023, hess_2024, Gonzalez-Valdes_Heydecker_Ortúzar_2022}). \citet{Gonzalez-Valdes_Heydecker_Ortúzar_2022} provide an example LCM using RUM, RRM, and elimination-by-aspects decision rules. Figure \ref{fig:latent_class_dc} provides the abstract structure, in which one uses (often logit) choice probabilities for inclusion in a latent class defining each decision rule heuristic. Importantly, different classes need not share a common utility-based structure. A class may form utilities, modify them, apply threshold or lexicographic rules, or employ heuristic processes that bypass utility maximization altogether. While similar in spirit to the modeling approach I will describe, LCMs do not provide a straightforward mechanism to disentangle the components of the vehicle choice. Figure \ref{fig:constraint_dc} shows the multi-constraint goal-based model that I will outline in this section.

As a starting point, consider an individual faced with a choice among several alternatives. They define thresholds that express self-imposed minimum requirements on value, which are \emph{soft} in the sense that violations (i.e., insufficiency of value) are subject to a penalty function. Let us further assume that the individual simultaneously pursues only two goals in the process of making their choice, each of which can be expressed as a separate optimization problem. Following the endogenous Maximum Entropy Program (eMEP) paradigm in \citet{Swait_Marley_2013}, we will assume the two goals correspond to expected utility maximization (an \emph{exploitation} goal) and maximizing choice set entropy (an \emph{exploration} goal). The simultaneous maximization of both goals is a vector optimization problem. The solution to such a problem is variously called \emph{non-inferior}, \emph{efficient}, \emph{Pareto optimal}, or \emph{non-dominated}. It can be solved by conversion into an equivalent scalar optimization problem. The conversion to a scalar problem makes the assumption that the individual is indifferent between maximizing expected value or choice set entropy. We can also replace the choice set entropy goal by a behavioral inertia function. Assume that the individual has prior information in the form of exogenous probabilities. The prior information may arise from prior experience, habits, or routine (repeated) behavior. Discrepancy between the two probability distributions measures adherence to past choice (sometimes termed \emph{relative entropy}), which can be quantified using a Kullback-Leibler information measure \citep{Swait_Marley_2013}.

The above model formulation moves us toward our ultimate goal of disentangling choice bundle attributes, but it assumes that a goal-balancing objective comprising linear combinations of those goals exists and that the goals are fully exchangeable across arguments of the sub-functions. This relationship may be difficult to represent in the presence of sub-function non-linearity. For example, a decision-maker may prefer foods with low salt, holding the taste of an item fixed, but they might be willing to tolerate a little more salt in a dish with extraordinary taste \citep{Hur_Allenby_2022}. Likewise, a decision-maker may not choose an alternative despite its high expected utility when they cannot tolerate even a small amount of anticipated regret. More fundamentally, multi-objective problems are formulated because it is difficult to conceive of collapsing the objectives onto a common scale. An alternative approach, which does not require the specification of a single objective function, is $\epsilon$-constraint optimization \citep{Haimes_Lasdon_Da_1971}. In the $\epsilon$-constraint method, all but one of the sub-objective functions are represented as constraints on the optimization of the remaining goal. \citet{Hur_Allenby_2022} provide a dual-goal implementation of $\epsilon$-constraint optimization, wherein the two goals are random utility maximization (RUM) and random regret minimization (RRM). They set the RUM goal as the objective and the RRM goal forms a constraint on the choice. Unlike the linear-combination-of-weights models proposed by \citet{Swait_Marley_2013}, this approach generates unique Pareto-optimal points by varying the upper bound on multiple constraints \citep{Chiandussi_Codegone_Ferrero_Varesio_2012}.

Using the multi-objective goal-based model framework of \citet{Hur_Allenby_2022}, we can consider a RUM goal of maximizing the benefits from the purchased vehicle at a minimum cost. The individual may be interested in the vehicle color, the number of seats, or other vehicle features that are relevant to their daily travel needs. Selecting an EV with a lower range will constrain their ability to complete long-distance travel, representing a RRM weighting of regret from not selecting an alternative vehicle with a longer range at a higher price point. Tying in environmental goals, the individual may be willing to trade-off driving range for a lower environmental impact relative to either an ICEV (from fossil fuel consumption) or long-range EV (from battery materials). We can further adapt the GPT theory literature by following the approach of \citet{Harman_Weinhardt_Gonzalez-Vallejo_Vancouver_2022}. They develop a survey instrument to capture affect and goal priming on decision search behavior. They examine accuracy and speed goals using word selection stimuli prior to a computer choice experiment. The approach could be adapted to vehicle choice using word stimuli associated with positive and negative affect regarding the above referenced goal---e.g., ``cramped", ``sleek", ``luxury" (vehicle goal); ``emissions", ``clean", ``tailpipe" (environmental goal); and ``freedom", ``waiting", ``range" (travel goal). Similarly, \citet{Steg_Perlaviciute_vanderWerff_Lurvink_2014} examine the relevance of hedonic goals to environmental attitudes and actions. They find that pro-environmental actions are influenced by hedonic goals, such as the desire to own a luxury vehicle. Such ``word stimuli`` could also represent mimetic cues by associating vehicle alternatives with particular social models or identities. For example, a survey might include images or descriptions of peers owning an EV, a luxury vehicle, or no personal vehicle to test whether exposure to an aspirational reference group changes the activation and weighting of environmental, status, and mobility goals. 

In summary, the $\epsilon$-constraint optimization approach allows us to assign goals to attributes of the vehicle and thus build an understanding of the partworth decision mechanisms.

\section{Application 3: Location Choice as a Matrix Factorization Problem}
Recommender systems are a common feature of e-commerce in a modern market of diverse products and heterogeneous consumer preferences. It is a large field of study that goes well beyond the scope of this paper. I focus on \emph{collaborative filtering}, a strategy that constructs user-item associations based on previous transactions or product ratings \citep{Koren_Rendle_Bell_2022}. An effective method to implement collaborative filtering is called \emph{matrix factorization} \citep{koren2009matrix}. This method defines a set of latent factors that characterize the decision-maker and the choice alternatives. Although the HCM framework is flexible enough to define latent variables for both decision-makers and alternatives, in practice latent variables are typically specified at the individual level and rely on psychometric indicators. Matrix factorization differs not in its theoretical capability but in its modeling strategy: it directly learns continuous latent representations for both individuals and alternatives from sparse interaction data, without relying on measurement models or a finite number of latent classes. This allows for the estimation of thousands or millions of latent factors without the need to explicitly specify measurement equations, structural equations, or class membership probabilities, as would be required in an HCM formulation. While less behaviorally interpretable, the matrix factorization approach has benefits in sparse respondent-alternative matrix applications, such as destination and home location choice problems.

\citet{Athey_Blei_Donnelly_Ruiz_Schmidt_2018} provide an empirical application of the above model to restaurant choice. Their matrix factorization model includes vectors representing individual latent preferences and alternative latent attributes, respectively. The model is specified as a hierarchical Bayesian logit model, wherein observed restaurant characteristics shift the distribution of the latent restaurant vector. The model is estimated using variational inference to reduce computational costs. The only prior reference in the transportation literature to the work of Athey et al. (of which I am aware) is by \citet{Krueger_Bierlaire_Daziano_Rashidi_Bansal_2021} in their work on inter- and intra-individual heterogeneity in mixed logit models. We can form a connection back to goal-based models through intra-individual heterogeneity by noting ``[u]nlike single goal research, the study of multiple goals takes seriously the intrapersonal dynamics of multiple goals to understand the operation of both single goals and the goals as a system" \citep[p. 3]{Kung_Scholer_2020}.

I consider the home location choice problem as an illustration of matrix factorization and its interpretation via GPT. The home location choice problem is generally framed from one of two perspectives. The first perspective (\emph{location choice}) is that an individual decides from among a choice set of available locations the one that maximizes their personal utility function \citep{Lerman_1976}. The second perspective is that multiple individuals bid on a location (\emph{bid rent}) and that home location alternative is allocated to the individual willing to offer the highest bid. \citet{Martinez_1992} proves the equivalence between these two approaches at equilibrium conditions.

\begin{figure*}[h]
  \centering
    \includegraphics[scale=0.4]{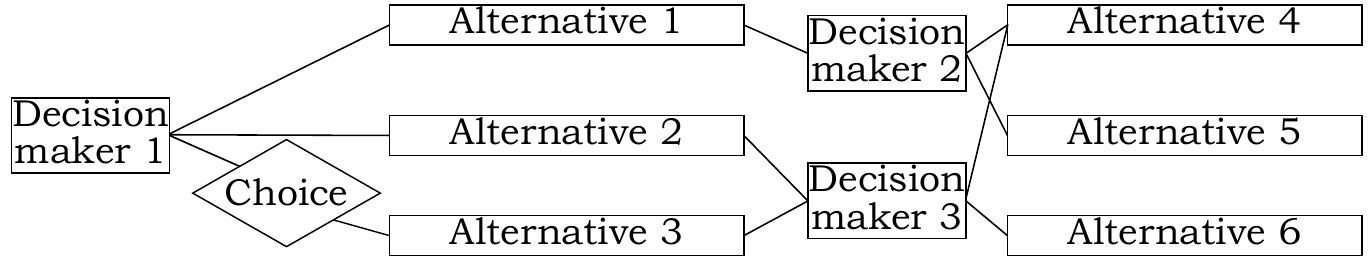}
  \caption{Matrix factorization discrete choice model structure} %
  \label{fig:matrix_factorization_dc}
\end{figure*}

An illustration using film choice provides a first step toward understanding the incorporation of matrix factorization into the home location choice problem. Recommender systems based on matrix factorization build a relationship between the films a person watches and unwatched films that may interest them based on the viewing patterns of similar individuals. In this context, \emph{similar is defined as individuals who also watched a particular film, while incorporating the other films watched by those individuals as additional information.} Let us now consider home location as replacing films in this example. The modeler can only observe a single home location choice for each household. The bid-choice equivalence suggests that similar households will also bid on this location, similar to the way that similar individuals will watch the same films. Furthermore, we can assume locations considered by similar households (but not the household of interest) may be of interest to them (i.e., in their choice set). Drawing a connection to GPT, households make tradeoffs between different goals when selecting a home. For example, a household may be interested in locating near a park, but it may be unable to find a suitable home near members' workplaces. In the same way that an individual may be observed to watch a horror film but have common interests with an individual observed to watch a comedy film, multiple goals in home location choice can be captured through latent variables that are common among households using a matrix factorization-based model formulation. The latent matrix relationships mirror the duality of bid-choice (see Figure \ref{fig:matrix_factorization_dc}). The key extension beyond standard LCMs offered by this matrix factorization model formulation is the associations made, not only between an individual decision-maker and their choice set, but also between that decision-maker and the chosen alternatives of similar decision-makers. In large choice set models, such as location choice problems, it is common that the choice set is unobserved and randomly assigned \citep{Hawkins_NurulHabib_2024}. Matrix factorization offers a mechanism to leverage the observed choices of similar decision-makers via dimensionality-reducing latent factors.

This application also illustrates the broader relationship between AI and behavioral theory developed in this paper. Matrix factorization can discover predictive latent relationships among households and locations, but those latent dimensions do not possess an inherent behavioral interpretation. GPT provides hypotheses about what these dimensions may represent---including goals, their instrumentality, and conflicts among them---that can subsequently be tested using observed indicators and behavioral validation.

\section{Operationalizing and Validating GPT in Transportation Research}
A recurring motivation of this discussion paper is that predictive fit and flexible functional forms are not substitutes for behavioral explanation. GPT is useful here because it forces the analyst to state (and test) a theory describing behavioral mechanisms: which goals are relevant, when they are activated, how conflicts are resolved, and how goal pursuit unfolds across time. This makes GPT a practical template for translating behavioral theory into empirical modeling choices, rather than treating theory as a post-hoc interpretation of a high-performing specification. 

Predictive performance should also be interpreted in relation to forecast uncertainty. \citet{Taleb_2009} argues that when forecast errors are difficult to estimate, decision-making should focus less on precise point predictions and more on identifiable properties of the modeled system, particularly its sensitivity to error and changing conditions. In transportation applications, this shifts attention from asking only whether a model accurately predicts a future behavior to asking whether its behavioral mechanisms and policy conclusions remain stable under plausible changes in goals, contexts, and
model assumptions. GPT is potentially useful from this perspective because it makes the choice-formation process explicit, allowing researchers to test which conclusions depend on fragile assumptions about goal activation and which remain robust across alternative behavioral conditions.

In the following, I outline a general process for GPT model development.

\subsection*{Step 1: Identify candidate goals for a specific decision context}
Because goals are latent, GPT applications require an explicit goal-construction step. A pragmatic approach is to begin with a parsimonious ``scaffold'' (e.g., hedonic, gain, and normative framings) and then propose context-specific goals that are (a) theoretically motivated and (b) decision-relevant. Goal identification can be strengthened by triangulating three sources of evidence: (1) prior behavioral theory and domain knowledge (what goals are plausible in this context?), (2) qualitative elicitation (interviews, focus groups, cognitive pretests), and (3) pilot survey evidence that the proposed goals are meaningfully activated and vary across individuals and situations. Importantly, goals can be represented at multiple levels of abstraction (e.g., ``reduce stress'' vs.\ ``avoid congestion''), so the analyst should justify the chosen level as appropriate to the temporal scale of the decision being modeled.

\subsection*{Step 2: Map goals to measurable indicators and choice mechanisms}
GPT becomes operational when hypothesized goals are linked to observable data via a measurement strategy and a mechanism within the choice process. Several complementary measurement approaches are available: (a) \emph{goal activation} indicators (binary or probabilistic activation in a specific choice situation); (b) \emph{goal importance} indicators (Likert scales or best--worst tasks measuring relative importance); (c) \emph{goal $\times$ alternative suitability/instrumentality} indicators (perceived extent to which an alternative can achieve a goal); (d) \emph{process measures} where feasible (response times, stated consideration sets); and (e) \emph{panel or multi-day diary data} to identify goal dynamics and adaptation across repeated decisions.

Mechanistically, goals may enter the model in at least three distinct ways: as (1) shifters of utility trade-offs (e.g., goal-weighted sensitivities to time, cost, or comfort), (2 drivers of consideration-set formation or screening (e.g., thresholds that exclude alternatives failing minimum goal attainment), and/or (3) constraints in multi-objective formulations (e.g., $\epsilon$-constraint approaches). Being explicit about which mechanism is claimed is critical: different mechanisms can yield similar fit but very different behavioral interpretations. AI methods may support this measurement process by extracting indicators of goal activation from interviews, open-ended survey responses, travel
diaries, or other unstructured data, and by estimating nonlinear relationships between context and goal activation. Such applications should remain theory-guided: computationally identified patterns become behaviorally meaningful only when linked to explicit constructs and testable mechanisms.

\subsection*{Step 3: Assess validity when goals are unobservable}
Because goals are latent, GPT should be accompanied by a transparent validation plan. Ideally, I recommend evaluating some combination of four forms of validity.

\textbf{Measurement validity.} If goals are measured by indicators, report reliability and basic measurement diagnostics (e.g., sensible ordering of thresholds in ordered-response models). Poor measurement can falsely suggest ``weak goals'' when the true issue is indicator noise.

\textbf{Construct validity.} Demonstrate that the proposed goals behave as the theory predicts. For example, goals expected to be distinct should exhibit discriminant validity; goals expected to co-activate under certain contexts should show corresponding correlations or activation patterns.

\textbf{Predictive validity and benchmarking.} Compare GPT-based specifications to clearly defined benchmarks (e.g., MNL/MXL, latent class, RRM, and/or flexible specifications) on the same dataset. Interpretability should be treated as a co-equal evaluation criterion: the goal-based structure should yield stable, meaningful behavioral narratives and policy-relevant mechanisms, not only improved likelihood metrics.

\textbf{Robustness and falsification.} Re-estimate with alternative goal sets, indicator mappings, activation
processes, and mechanisms (screening vs.\ constraints). Following \citet{Taleb_2009}, evaluation should examine not only expected predictive error but also whether small modeling errors or contextual changes produce disproportionately large changes in forecasts or policy conclusions. Whenever possible, test stability across subpopulations and across time (panel data). These exercises reduce researcher degrees of freedom and help guard against ``goal overfitting.''

\subsection*{Step 4: Anticipate practical limitations}
GPT is empirically demanding. Additional survey components increase respondent burden and raise identification concerns; richer latent structures increase estimation complexity; and goals may evolve with life events, habituation, or learning, complicating forecasting and transferability. These limitations do not negate GPT's value, but they imply that GPT applications should be explicit about (a) which goals are included and why, (b) what evidence supports the measurement strategy, and (c) what aspects of the model are intended to be explanatory rather than purely predictive.

\subsection*{Summary checklist}
In operational terms, a GPT application should be able to answer the following questions:
\begin{enumerate}
    \item \textbf{Goal set:} What goals are hypothesized for this decision context, and what is the rationale for inclusion or exclusion?
    \item \textbf{Mechanism:} Do goals affect utility trade-offs, consideration or screening, constraints, or sequential decisions (and why)?
    \item \textbf{Measurement:} What indicators capture activation, importance, and instrumentality?
    \item \textbf{Validation:} What benchmark models are used, and what evidence supports construct and predictive validity?
    \item \textbf{Robustness:} How sensitive are results to alternative goal sets, subpopulations, and temporal dynamics?
\end{enumerate}

\section{Conclusions and Future Work}
This paper has used goal pursuit theory (GPT) as an illustrative example of why travel behavior research benefits from behavioral theory, not only flexible statistical specifications. GPT addresses key theoretical limitations of traditional RUM and RRM frameworks by making the underlying motivational structure explicit. Through three applications---activity scheduling, vehicle ownership, and location choice---I showed how GPT can model (1) context-dependent goal activation, (2) resolution of competing objectives, and (3) multi-temporal decision processes.

A central implication is that ``better fit'' is not synonymous with ``better understanding''. Increasingly flexible functional forms can improve predictive performance, but without an explicit theory of motivation one risks producing models that are difficult to interpret, diagnose, or translate into policy-relevant mechanisms. GPT provides one concrete route to re-centering explanation by tying observed choices to latent goals, their activation, and the trade-offs that unfold across situations and time.

Returning to the motivating question of artificial intelligence, our argument is not that travel behavior researchers must choose between AI and behavioral theory. AI methods can flexibly represent nonlinear relationships, identify behavioral heterogeneity, process large and unstructured datasets, and improve predictive performance. However, these capabilities do not independently establish which goals matter, how they are activated, whether estimated associations represent
behavioral mechanisms, or how behavior will respond when policies and contexts change. GPT illustrates a complementary division of labor: behavioral theory defines the relevant constructs, mechanisms, and falsifiable expectations, while AI provides tools for measurement, estimation, pattern discovery, and testing at scale. From this perspective, AI is most valuable not as a replacement for travel behavior theory, but as a means of operationalizing, evaluating, and
refining it.

Our hope is that the conceptual frameworks discussed herein spur empirical applications and comparisons against alternative decision theories and model formulations, with evaluation not limited to prediction accuracy but also interpretability and behavioral plausibility. Empirical research will require innovative survey instruments that capture both functional and higher-level latent goal constructs. Multi-day and panel travel diary collection also offer avenues to better understand goal dynamics in everyday travel. Furthermore, \citet{Chartrand_Bargh_1999} identify unconscious goal pursuit through external \emph{goal priming} (e.g., social interactions), which may offer a connection between GPT and the recent focus in the travel behavior research field on the adoption of behavioral science methods to establish innovative policy mechanisms.

The appropriate model structure among goal-based models will depend on behavioral considerations, including constraints on choice and the research question(s) under study. Much of the research to date has focused on conceptual GPT development or econometric goal-based model specification; there remains a need for cumulative empirical work that clarifies when, where, and for whom goal-based theories provide distinct insights beyond conventional utility specifications. A key open question is how goals vary as a function of sociodemographics and evolve across life stages. These questions, in turn, raise issues of transferability and forecasting when goals themselves change over time.

GPT also provides guidance for policy development, as it emphasizes the interplay between functional and normative goals. For example, transport policies promoting sustainable travel may be more effective when they consider both personal and societal motivations, and when they target mechanisms that facilitate (or prime) the activation of pro-social goals in relevant contexts. Limitations of the present work include its conceptual focus and lack of empirical validation. Future research should apply GPT in practical contexts, test its predictive and explanatory capabilities against benchmark RUM/RRM specifications, and explore its integration with large-scale transport demand models to inform evidence-based policy decisions.

A final point on GPT is that work to date has largely focused on individual goal framings. My most recent thoughts on the subject have been influenced by \cite{Sagoff_2008}, who distinguishes between two social roles we play: the citizen and the consumer. Econometric models predominantly frame choice from the consumer perspective, meaning consideration of a personal utility function. However, in political contexts, it is often the case that we make decisions as citizens considering public welfare. There is much to be said on this topic. For now, suffice to note that community-oriented GPT could be a fruitful direction that extends standard individualistic utility with emerging work in social network representations.

\section*{Acknowledgments}
The first author is grateful to Omid Armantalab for assistance with the literature review and insightful feedback on the paper. Thanks also to Joffre Swait for his insights on goal-based choice models. The encouragement of Byron Miller, Eric Miller, and Khandker Nurul Habib to pursue this line of theoretical research is also much appreciated.

\bibliography{references}

\newpage
\appendix
\appendixpage
These appendices provide the mathematical details for the models discussed in the body of the paper. I provide notation in a table for each model formulation. 

\section{Application 1: Activity Scheduling as a Hierarchical and Sequential Goal Problem}
The full goal-based HCM model structure is illustrated in Figure \ref{fig:latent_structure} as a conceptual model, components of which have been implemented in various forms \citep{Marley_Swait_2017, Swait_Franceschinis_Thiene_2020}. Capacity limits can be implemented in the model through scale parameter heterogeneity, which is interpreted as stemming from individual characteristics, including cognitive capacity constraints. Note that cognitive capacity effects in the goal-based HCM model parallel the motivation for rational inattention \citep{Habib_2023} and there may be opportunities to incorporate the two concepts within an econometric model structure.

\subsection{HCM Formulations}
The hybrid choice model (HCM) of \citet{Swait_Franceschinis_Thiene_2020} is decomposed into structural indicator and discrete choice components, making it a form of integrated choice and latent variable (ICLV) model. Notation for this model is provided in Table \ref{tab:iclv_parameters}.

\begin{table}[h]
  \caption{Parameters and variables of the HCM}
  \label{tab:iclv_parameters}
  \centering
\begin{tabular}{cl}
\toprule
$w$ & Latent importance indictor for goals          \\
$G$ & Binary activation function for goals \\
$I$ & Measured importance indicator for goals \\
$\chi$ & \makecell[l]{Vector of constants for the observed answers to the goal activation questions} \\ 
$\vartheta$ & \makecell[l]{Vector of coefficients for the goal importance latent variables for the observed answers\\to the goal activation questions}\\
$\xi$ & \makecell[l]{Propensity to consider a choice to be sufficiently suitable to achieve a goal} \\
$\mu$ & \makecell[l]{Vector of coefficients for the observed scores on Likert scales measuring goal importance\\as result of the goal importance latent variables}   \\ 
$\tau$ & \makecell[l]{Vector of thresholds explaining the observed scores on the Likert scales measuring\\goal importance}\\
$H$ & Vector of constants for the observed answers to the goal $\times$ choice question                  \\
$\omega$ & \makecell[l]{Vector of coefficients explaining the answers to the goal $\times$ choice question\\as result of the goal $\times$ choice  latent variables} \\
$\beta$ & Vector of coefficients explaining the weight of choice attributes on the utility function                        \\
$\phi$ & Vector of coefficients explaining the impedance (distance) on the utility function                               \\
$\theta$ & Vector of coefficients explaining the weight of goal $\times$ choice propensity                                         \\
$\gamma$ & Vector of coefficients measuring the fixed effect of impedance (distance) on $\phi$                                             \\
$\delta$ & Vector of coefficients explaining the effect of latent goal importances
on $\rho$ \\
$\alpha$ & Vectors of parameters associated with latent goal importances in the h() function \\
$\lambda$ & \makecell[l]{Vector of parameters scaling the effect of $\theta$ on utility as function of latent goal importances \\ through f() function}                \\                               
$\pi$ & Vector of constants for the class allocation model \\
\bottomrule
\end{tabular}
\end{table}

Indicators in the HCM model are given by structural equation estimates for latent variable, $w_g$, which measure the importance of goal $g$ through two sets of indicators: $G_g$ is a binary activation function and $I_g$ is an indicator of the importance of the goal. We can ignore the decision-maker index for simplicity. The activation function is measured by the following equation, with $\chi_g$ being a goal-specific constant and $\vartheta_g$ measuring the impact of latent variable $w_g$ on the probability of pursuing goal $g$.

\begin{equation}
    P(G_g=1|\chi, \vartheta,w_g) = \frac{\exp(\chi_g + \vartheta_g w_g)}{1+\exp(\chi_g + \vartheta_g w_g)}
\end{equation}
The measurement equation for the second set of indicators is given by an ordered process model for a latent scale as follows
\begin{equation}
    P(I_g=l_k|w_g) = \frac{\exp(\tau_k + \mu_g w_g)}{1+\exp(\tau_k - \mu_g w_g)} - \frac{\exp(\tau_{k-1} - \mu_g w_g)}{1+\exp(\tau_{k-1} - \mu_g w_g)}
\end{equation}
where $\tau_1, \cdots, \tau_{L-1}$ are thresholds.

The $I_g$ indicator function is given as follows
\[
I_g = k \quad \text{if} \quad \tau_{k-1} < w_g \le \tau_k
\]

The propensity to consider choice $k$ to be sufficiently suitable to achieve goal $g$ is measured by a second group of latent indicator variables $\xi_{gk}$. The $GK_{gk}$ indicators are equal to 1 if choice $k$ is considered suitable to achieve goal $g$, giving a logistic regression. The logistic model is given by
\begin{equation}
    P_{GK_{gk}}(H,\theta|\xi_{gk}) = \frac{\exp(H_k + \omega_k \xi_{gk})}{1+\exp(H_k + \omega_k \xi_{gk})}
\end{equation}
where $H_k$ is a constant and $\omega_k$ measures the impact of $\xi_{gk}$ on the probability of choice $k$ being suitable to achieve goal $g$. In this model, $\xi_{gk}$ is an indicator measured on a latent scale for a given choice $k$ and goal $g$ taken from a survey question. 

The probability is measured using random utility maximization (RUM). The utility is given by
\begin{equation}
    U_k = ASC_k + \beta X_k + \sum_{g=1}^G \theta_{gk}f(w_g)\xi_{gk} + \epsilon_k
\end{equation}
where $X_k$ is a set of covariates for alternative $k$ and with
\begin{equation}
    f(w_g) = \frac{1}{1+\exp(-\lambda_g w_g)}
\end{equation}
where $\lambda_g$ is a scale factor. The better choice $k$ is considered to satisfy goal $g$, the higher $\xi_{gk}$ will be. Likewise, the more important a goal is considered to be, the closer $f()$ will be to unity so that the impact of $\xi_{gk}$ is strengthened. The unobserved error $\epsilon_k$ is assumed to be i.i.d. extreme value type I.

\citet{Swait_Franceschinis_Thiene_2020} use a latent class approach to capture heterogeneity in $P_{nc}$, the choice probability for class $c$ and individual $n$, where $\pi_c$ is a class-specific constant term (as defined by \citet{Scarpa_Thiene_2005}) and given by an MNL formulation and $\pi_1 = 1$ for identification.

The impedance cost to access a choice is a function of goal importance and has coefficient $\rho_k$ given by
\begin{equation}
    \rho_k = \gamma_k + \sum_{g=1}^G \delta_{gk} h(w_g)
\end{equation}
where $\gamma_k$ is the fixed effect of impedance cost, $\delta_g$ measures the impact of goal $g$ on $\rho_k$, and
\begin{equation}
    h(w_g) = \frac{1}{1+\exp(-\alpha_g w_g)}
\end{equation}
where $\alpha_g$ is a scale factor.

Finally, the choice probability of alternative $k$ from the set of alternatives $J$ in choice occasion $t$ conditional on class $c$ is given by an MNL logit formula.

\begin{equation}
    P_{ktc}(\beta, \mu, \rho|\xi_n) = \frac{\exp(ASC_{kc} + \beta_{kc}X_k + \sum_{g=1}^G \theta_{gk}f(w_g)\xi_{gk})}{\sum_{j=1}^J \exp(ASC_{jc} + \beta_{jc}X_k + \sum_{g=1}^G \theta_{gj}f(w_g)\xi_{gj})}
\end{equation}

The full likelihood function is given by
\begin{equation}
    \begin{split}
        LL_N & = \sum_{n=1}^N (\ln(\sum_{c=1}^C (\prod_{t=1}^T P_{k_t^*tc}(\beta_c,\mu,\rho|\xi_{k_t^*}))P_{nc})) \\
        & +\sum_{g=1}^G(E(G_g=1)\ln(P(G_g=1|\chi, \vartheta,w_g))+E(G_g=0)\ln(1-P_{Gg}(K,\vartheta|w_g))) \\
        & +\sum_{g=1}^G\sum_{k=1}^K(E(GK_{gk}=1)\ln(P_{GK_{gk}}(H,\theta|\xi_{gk}))+E(GK_{gk}=0)\ln(1-P_{GK_{gk}}(H,\theta|\xi_{gk}))) \\
        & +\sum_{g=1}^G\sum_{l=1}^L(E(I_g=l)\ln(P(I_g=l|w_g)))))
    \end{split}
\end{equation}
where $E(x)$ equals 1 if event $x$ is true and 0 otherwise; $t, T$ are task indices; $k_t^*$ is the choice for task $t$; and other quantities are as defined above.

\subsection{2-stage Choice Formulation}
\citet{Marley_Swait_2017} provide a 2-stage goal-based choice model as a hierarchical decision-making process wherein long-term abstract goals are dealt with in a choice set formation stage, following which choice-stage functional goals are then pursued
within the choice stage. The model implements a \emph{hierarchical} GPT structure, though the authors do not make an explicit reference as such. The first stage comprises a goal-based choice set formation process for a set of goals $\mathcal{A}$ and strategy $\Phi_{\mathcal{M}}(\mathcal{A})$. The second stage comprises a choice from among the activated $\mathcal{C} \in \Gamma(\mathcal{M})$ goals based on a set of goals $\mathcal{B}$ and goal attainment strategy $\Psi_{\mathcal{C}}(\mathcal{B})$. The goal set $\mathcal{A}$ (and strategy $\Theta(\mathcal{A})$) can be different than the goal set $\mathcal{B}$ (and strategy $\Psi_{\mathcal{B}}(\mathcal{B})$). Notation for this model is provided in Table \ref{tab:two_stage_parameters}.

\begin{table}[hb]
  \caption{Parameters and variables of the two-stage model}
  \label{tab:two_stage_parameters}
  \centering
\begin{tabular}{cl}
\toprule
z & An alternative $\in \mathcal{C}$ \\
$\mathcal{A}$ & Set of goals \\
$\mathcal{B}$ & Set of goals \\
$\mathcal{C}$ & Set of activated alternatives \\
$\mathcal{M}$ & Set of available alternatives \\
$\Gamma(\mathcal{\cdot})$ & Set of non-empty subsets for variable of interest \\
$\mathcal{G}$ & Set of currently activated functional goals \\
$\Theta_{\mathcal{M}}(\mathcal{A})$ & Goal attainment strategy \\
$\Psi_{\mathcal{C}}(\mathcal{B})$ & Goal attainment strategy \\
$\mu_{g\mathcal{M}}$ & Scale parameters for a goal $g$ and choice set $\mathcal{M}$ \\
$\sigma_{h\mathcal{C}}$ & Scale parameters for a goal $h$ and choice set $\mathcal{C}$ \\
$\S_g\mathcal{C}$ & Utility for choice set $\mathcal{C}$ \\
$V_{h\mathcal{C}}(z)$ & Utility for alternative $z$ \\
$\tau_g$ & Threshold defining minimum level for goal $g$ \\
$\alpha$/$\lambda$ & Parameters in Kahneman and Tversky threshold function \\  
\bottomrule
\end{tabular}
\end{table}

\textbf{Choice set formation:} With $\mathcal{M}$ being the available choice set, $a_M(\mathcal{A})$ is the probability that the set of goals $\mathcal{A} \in \Gamma(\mathcal{G})$ is activated and $q_{\Gamma(M)}(C|\Theta_{\mathcal{M}}(\mathcal{A}))$ is the probability of (sub)set $C \in \Gamma$ being chosen under goal strategy $\Theta_M(A)$.

\textbf{Within-set choice:} With $\mathcal{C}$ being the focal subset, $b_{\mathcal{C}}(\mathcal{B})$, is the probability of set of goals $\mathcal{B} \in \Gamma(\mathcal{G})$ being activated and $r_{\mathcal{C}}(z|\Psi_{\mathcal{C}}(\mathcal{B}))$ is the probability of alternative $z \in \mathcal{C}$ being chosen under goal strategy $\Psi_{\mathcal{C}}(\mathcal{B})$. $P_{\Gamma(\mathcal{G})_{\mathcal{M}}}(z)$ is the overall probability of selecting $z$ from choice set $\mathcal{M}$. The formulation is then given by the following

\begin{equation}
\label{eq:two-stage}
P_{\Gamma(\mathcal{G})_\mathcal{M}}(z) = \sum_{C \in \Gamma (\mathcal{M})}\left(\sum_{\mathcal{A} \in \Gamma(\mathcal{G})} a_\mathcal{M}(\mathcal{A})q_{\Gamma(\mathcal{M})}(\mathcal{C}|\Theta_{\mathcal{M}}(\mathcal{A})\right)\left(\sum_{\mathcal{B} \in \Gamma(\mathcal{G})} b_c(\mathcal{B})r_{\mathcal{C}}(z|\Psi_{\mathcal{C}}(\mathcal{B})\right)
\end{equation}
Including psychological variable such as values, beliefs, and attitudes (omitted above) gives the following full equation
\begin{equation}
P_{\xi \Gamma(\mathcal{G})_\mathcal{M}}(z) = \sum_{C \in \Gamma (\mathcal{M})}\left(\sum_{\mathcal{A} \in \Gamma(\mathcal{G})} a_{\xi M}(A)q_{\xi\Gamma(\mathcal{M})}(\mathcal{C}|\Theta_{\mathcal{M}}(\mathcal{A})\right)\left(\sum_{\mathcal{B} \in \Gamma(\mathcal{G})} b_{\xi c}(\mathcal{B})r_{\xi \mathcal{C}}(z|\Psi_{\mathcal{C}}(\mathcal{B})\right)
\end{equation}
These variables become relevant when an individual has a goal to associate with people with shard beliefs and values, or in other similar decision contexts.

One can define the following probability terms from Equation \ref{eq:two-stage}

\begin{equation}
Q_{\Gamma(\mathcal{G})\Gamma(\mathcal{M})}(\mathcal{C}) = \sum_{\mathcal{A} \in \Gamma(\mathcal{G})} a_{\mathcal{M}}(\mathcal{A}) q_{\Gamma(\mathcal{M})}(\mathcal{C}|\Theta_{\mathcal{M}}(\mathcal{A}))
\end{equation}
and
\begin{equation}
R_{\Gamma(\mathcal{G})\mathcal{C}}(z) = \sum_{\mathcal{B} \in \Gamma(\mathcal{G})} b_c(\mathcal{B})r_c(z|\Psi_{\mathcal{C}}(\mathcal{B}))
\end{equation}
The following simplified form results for Equation \ref{eq:two-stage}
\begin{equation}
P_{\Gamma(\mathcal{G})\mathcal{M}}(z) = \sum_{c \in \Gamma(\mathcal{M})} Q_{\Gamma (\mathcal{G}) \Gamma (\mathcal{M})}(\mathcal{C})R_{\Gamma(\mathcal{G})\mathcal{C}}(z)
\end{equation}

The above is a standard two-stage model, with $Q_{\Gamma (\mathcal{G}) \Gamma (\mathcal{M})}(\mathcal{C})$ being the choice set model and $R_{\Gamma(\mathcal{G})\mathcal{C}}(z)$ being the choice model. The modeller will often ignore $\Gamma(G)$ in a standard two-stage model because it would simply represent RUM (or RRM in some recent applications). In the below, I replace $\mathcal{G}$ with $g$. I can then define the following probability function, which is a goal-based extension of set-dependent MNL (see \cite{Marley_Flynn_Louviere_2008}):
\begin{equation}
P_{g\mathcal{M}}(z) = \sum_{\mathcal{C} \in \Gamma(\mathcal{M})} Q_{g \Gamma(\mathcal{M})} (\mathcal{C}) R_{h\mathcal{C}}(z)
\end{equation}
with
\begin{equation}
Q_{g \Gamma(\mathcal{M})} (\mathcal{C}) = \frac{\exp(\mu_{g\mathcal{M}}S_g(\mathcal{C})}{\sum_{\mathcal{D} \in \Gamma(\mathcal{M})} \exp(\mu_{g\mathcal{M}}S_g(\mathcal{C})}
\end{equation}
and
\begin{equation}
R_{h\mathcal{C}} (z) = \frac{\exp(\sigma_{h\mathcal{C}}V_{hC}(z))}{\sum_{s \in \mathcal{C}} \exp(\sigma_{s\mathcal{C}}V_{s\mathcal{C}}(z))}
\end{equation}
where each $\mu_{gM}$ and $\sigma_{hC}$ is nonnegative. The above representation of the model can be motivated by the endogenous Maximum Entropy Program (eMEP) paradigm in \citet{Swait_Marley_2013}, which captures stochasticity as a consequence of the decision-maker having multiple goals that cannot be simultaneously achieved at their maximal levels. Maximizing expected utility at each level is an \emph{exploitation} goal and maximizing entropy is an \emph{exploration} goal.

\subsubsection{Goal Thresholds as a Screening Mechanism}
One can consider goal thresholds as a screening mechanism in the choice set stage. With an available set of alternatives $\mathcal{M}$, alternative $z$, and set of goals $\mathcal{G}$ with $g$ being a typical goal in the set, let $V_g(z)$ be the attainability of goal $g$ by alternative $z$ and $\tau_g$ be a desired minimum value level for goal attainability. Then let $f(V_g(z)-\tau_g)$, with $f(0)=0$ being nondecreasing and nonlinear, be a measure of alternative accessibility relative to the threshold $\tau_g$. Let $f(0)=0$ be a measure of how acceptable alternative $z$ is relative to threshold $\tau_g$. The \citet{Kahneman_Tversky_1979} threshold function that implies deviations below the threshold are weighted more heavily than deviations above it is given by 
\begin{equation}
    f(r)= 
\begin{cases}
    r^{\alpha}& \text{if } r\ge 0\\
    -\lambda(-r)^{\alpha}              & \text{otherwise}
\end{cases}
\end{equation}
The probability that alternative $z$ is considered acceptable is then given by the following logistic function
\begin{equation}
P_{g\mathcal{M}}(z) = \frac{1}{1+\exp(-\sigma_{g\mathcal{M}}f(V_g(z)-\tau_g)}
\end{equation}
where $\sigma_{g\mathcal{M}} > 0$ and is an expression of the difficulty of the task and/or the importance of the goal $g$ to the decision-maker. Other model details are provided in \citet{Marley_Swait_2017}.

\section{Application 2: Disentangling Vehicle Choice}
\subsection{eMEP constrained Optimization Formulation}
The first optimization approach continues from the prior model, with a focus on the eMEP approach from \citet{Swait_Marley_2013}. Notation for this model (in additional that defined in Table \ref{tab:two_stage_parameters}) is provided in Table \ref{tab:emep_parameters}.

\begin{table}[h]
  \caption{Parameters and variables of the eMEP constrained optimization model (in addition to those in Table \ref{tab:two_stage_parameters})}
  \label{tab:emep_parameters}
  \centering
\begin{tabular}{cl}
\toprule
$\tau_{i\mathcal{C}}$ & Threshold for minimum utility value associated with pursuit of a goal for alternative $i$ \\
$\kappa_{i\mathcal{C}}$ & Penalty on violation of the threshold $\tau_{i\mathcal{C}}$ for alternative $i$  \\
$w_{\mathcal{C}}$ & Scalar goal combination weight  \\
\bottomrule
\end{tabular}
\end{table}

Consider an individual faced with a choice set $\mathcal{C}$. They assign a deterministic value $V_j$ to each alternative $j \in \mathcal{C}$. There are also thresholds $\tau_{j\mathcal{C}}$ that express self-imposed minimum requirements on value, which are \emph{soft} in the sense that violations (i.e., insufficiency of value) are penalized by $\kappa_{k\mathcal{C}}$ value units per unit of violation. Hence, the effective utility of an alternative is given by $\kappa_{j\mathcal{C}}(V_j - \tau_{j\mathcal{C}})$. Let us assume that the individual simultaneously pursues only two goals in the process of making their choice, each of which can be expressed as a separate optimization problem. Furthermore, we can assume the two goals correspond to expected value maximizing consumption (\emph{exploitation}) and variety seeking choice set entropy (\emph{exploration}). The first goal can be formulated as an expected value of utility as follows
\begin{equation}
    \begin{split}
    \text{max } U(P, \mathcal{C}) =& \sum_{i \in \mathcal{C}} P_{i\mathcal{C}}\kappa_{i\mathcal{C}}(V_i - \tau_{iC}) \\
    s.t. \sum_{i \in \mathcal{C}} P_{i\mathcal{C}} =& 1\text{, } P_{i\mathcal{C}} \ge 0 \text{, } \forall i \in C
    \end{split}
\end{equation}
where $P_{i\mathcal{C}}$ is the choice probability for alternative $i$ in choice set $\mathcal{C}$. The above optimization can be solved in isolation as a standard logit model. 

The second goal (choice set entropy goal) seeks to maximize the information entropy content (i.e., choice set variety) by solving the following problem
\begin{equation}
    \begin{split}
    \text{max }\Psi(P, \mathcal{C}) =& - \sum_{i \in \mathcal{C}} P_{iC} \ln(P_{i\mathcal{C}}) \\
    s.t. \sum_{i \in \mathcal{C}} P_{i\mathcal{C}} =& 1\text{, } P_{i\mathcal{C}} \ge 0 \text{, } \forall i \in \mathcal{C} \\
    \end{split}
\end{equation}

The simultaneous maximization of both goals is a vector optimization problem. The solution to such a problem is variously called \emph{non-inferior}, \emph{efficient}, \emph{Pareto optimal}, or \emph{non-dominated}. The vector optimization problem can be solved by conversion into an equivalent scalar optimization problem. The conversion to a scalar problem makes the assumption that the decision-maker is indifferent between maximizing expected value or choice set entropy, and the decision-maker seeks to maximize choice probabilities that meet non-negativity requirements and add up to one over the set $\mathcal{C}$. 

The scalar optimization problem is taken as the convex combination of the expected value and choice set entropy goals defined above, with $w_{\mathcal{C}} \in[0,1]$
\begin{equation}
    \label{eq:max_function}
    \begin{split}
    \text{max } \Tilde{\Delta}(P, \mathcal{C}) =& w_{\mathcal{C}} U(P,\mathcal{C}) + (1-w_{\mathcal{C}})\Psi(P,\mathcal{C}) \\
    =& w_{\mathcal{C}} \sum_{i \in \mathcal{C}}(P_{i\mathcal{C}}\kappa_{i\mathcal{C}}(V_i - \tau_{i\mathcal{C}}) \\
    & -(1-w_\mathcal{C})\sum_{i \in \mathcal{C}}P_{i\mathcal{C}}\ln(P_{i\mathcal{C}}) \\
    s.t. \sum_{i \in \mathcal{C}}P_{i\mathcal{C}}=& 1 \text{, } P_{i\mathcal{C}} \ge 0 \text{, } \forall i \in C
    \end{split}
\end{equation}
The problem can be restated as follows
\begin{equation}
    \text{max } \Delta(P, \mathcal{C}) = r_{\mathcal{C}} U(P,\mathcal{C}) + \Psi(P,\mathcal{C})
\end{equation}
where $r_{\mathcal{C}} = w_{\mathcal{C}}/(1-w_{\mathcal{C}})$.

Finally, the choice probability that solves the above problem is given by one of three cases, determined by the value of parameter $r_{\mathcal{C}}$

\begin{equation*}
    \begin{split}
    \text{Case 1: }& r_{\mathcal{C}} = 0 \text{, equivalently } w_{\mathcal{C}} = 0  \\
    & P_{j\mathcal{C}} = J^{-1}, \forall j \in \mathcal{C}, J = |C| \\
    \text{Case 2: }& r_{\mathcal{C}} > 0 \text{ and finite, equivalently } w_{\mathcal{C}} \neq 0, 1 \\
    &P_{iC} = \frac{\exp(r_{\mathcal{C}} \kappa_{i\mathcal{C}}(V_i - \tau_{i\mathcal{C}})}{\sum_{J \in \mathcal{C}} \exp(r_C \kappa_{j\mathcal{C}}(V_j - \tau_{j\mathcal{C}})},  \forall j \in C \\
    \text{Case 3: }& r_{\mathcal{C}} > \infty \text{, equivalently } w_{\mathcal{C}} = 1  \\
    &M_C =  \argmax_{k \in \mathcal{C}}(\kappa_{kC}(V_k - \tau_{k\mathcal{C}})) \\
    \text{(a) } & \text{If the cardinality} |M_{\mathcal{C}}| = 1 \text{, the optimal choice probabilities are} \\
    &P_{j\mathcal{C}} =
    \begin{cases}
        1 & \text{if } j\in M_{\mathcal{C}} \\
        0 & \text{otherwise}
    \end{cases} \\
    \text{(b) } & \text{If the cardinality} |M_{\mathcal{C}}|>1\text{, any probability distribution over } M_{\mathcal{C}} \\
    & \text{that satisfies} \\
    & \sum_{i \in M_{\mathcal{C}}} P_{i\mathcal{C}} = 1\text{, } P_{i\mathcal{C}} \ge 0\ \forall i \in M_{\mathcal{C}} \\
    & \qquad P_{i\mathcal{C}} = 0\ \forall i \in (\mathcal{C} \setminus M_{\mathcal{C}}) \\
    & \text{will maximize Equation \ref{eq:max_function}}
    \end{split}
\end{equation*}
It can be shown that the above result is a concave and unique solution.

\subsection{$\epsilon$-Constraint Optimization Formulation}
The formulation of the prior subsection uses an overall objective that combines a set of sub-objectives with appropriate weights. This formulation assumes that a goal-balancing objective comprising linear combinations of those goals exists and that the goals are fully exchangeable across arguments of the sub-functions. An alternative approach, which does not require the specification of a single objective function, is known as the $\epsilon$-constraint method \citep{Haimes_Lasdon_Da_1971}. In the $\epsilon$-constraint method, all but one of the sub-objective functions are represented as constraints to the optimization of the remaining goal. There are $N$ possible formulations in this approach for $N$ goals, as each can be selected as the remaining goal. It is therefore necessary to compare which formulation fits the observed data best. Notation for this model is provided in Table \ref{tab:two_goal_model}.

\begin{table}[h]
  \caption{Parameters of the $\epsilon$-constraint optimization model}
  \label{tab:two_goal_model}
  \centering
\begin{tabular}{cl}
\toprule
$R_i$ & Random regret function for alternative $i$ \\
$U_i$ & Random utility function for alternative $i$ \\
$V_i$ & Deterministic utility for alternative $i$ \\
$\eta$ & Error term \\
$\bm{p}$ & Price vector \\
$\bm{x}$ & Consumption vector \\
$E$ & Total expenditure \\
\bottomrule
\end{tabular}
\end{table}

\citet{Hur_Allenby_2022} provide a dual-goal implementation of the $\epsilon$-constraint formulation, wherein the two goals are random utility maximization (RUM, $U(x)$) and random regret minimization (RRM, $R(x)$).  They include an additional uncertainty term (beyond the standard error associated with uncertainty by the researcher) to capture choice uncertainty by the decision-maker. I do not consider this innovation in my review of the formulation.

Defining utility associated with alternative $i$ as $U_i$, the regret function is defined by
\begin{equation}
R_i = \max_{j \neq i} (U_j) - U_i
\end{equation}

The dual objective function is then given by
\begin{equation}
    \begin{split}
    \max U(x) & \text{ } \min R(x) \\
    \text{s.t. } \bm{p}'\bm{x} &\leq E
    \end{split}
\end{equation}

where $E$ is the budget, $\bm{x}$ is a vector of alternatives, and $\bm{p}$ is its associated vector of prices.

Setting RUM as the objective, with RRM forming an additional constraint, the objective function is given by the following (expenditure $E$ cancels out for the singly discrete choice case)
\begin{equation}
    \label{eq:dual_goal}
    \begin{split}
    \max & U(x) \\
    \text{s.t. }& R(x) \le \theta
    \end{split}
\end{equation}
where $\theta$ is an endogenously determined threshold parameter. Alternatives whose anticipated regret exceed the threshold are screened out of the decision-makers' consideration set. This method solves the dual-goal problem and produces a unique solution. However, it may be the case that the optimal solution does not maximize both objectives - i.e., a decision-maker may maximize one goal and settle for the other goal achieving an acceptable level. \citet{Hur_Allenby_2022} use a two-stage choice model wherein RRM dictates the choice set process, which forms a constrain on a RUM choice model. The full likelihood for the choice of alternative $i_m$ is given below
\begin{equation}
    P(i_m) = P(i_1, \cdots, i_m \in C) P(i_m| i_1, \cdots, i_m \in C)
\end{equation}

The first part of the equation, $P(i_1, \cdots, i_m \in C)$, measures the probability that alternatives $m \in C$ are included in the choice set, as shown below
\begin{equation}
    \label{eq:first_stage}
    \begin{split}
    P(i_1, \cdots i_m \in C) = & P (R_{i_1} \le \theta, \cdots R_{i_m} \le \theta) \\
    & = P \left( \sum_{n \neq i_1} (V_n + \eta_n) \prod_{m \neq n} (V_n + \eta_n - V_m - \eta_m) - V_{i_1} - n_{i_1} \leq \theta, \cdots, \right. \\
    & \left. \sum_{n \neq i_m}(V_n + \eta_n)\prod_{m \neq n} (V_n + \eta_n - V_m - \eta_m) - V_{i_m} - n_{i_m} \leq \theta \right)
    \end{split}
\end{equation}
Note that the joint probability above cannot be factored into the product of marginal distributions, as the error term $\eta$ of one alternative affects the anticipated regret of the other alternatives. The second part of Equation \ref{eq:dual_goal} is given below

\begin{equation}
    \begin{split}
    P(i_m|i_1, \cdots, i_m \in C) = & P(V_{i_m} + \eta_{i_m} > V_{i_n} \\
    &+ \eta_{i_n}\text{, } \forall n \neq m|i_m, i_n \in C, \eta_{i_m}\text{, } \eta_{i_n} \in \Xi)
    \end{split}
\end{equation}
where $\Xi = {\eta|R_i \leq \theta \text{, } \forall i \in C}$. As the final choice is among alternatives from an endogenous choice set model, the choice is conditional upon the error term $\eta$ and the first stage model given in Equation \ref{eq:first_stage}. The model is estimated as a joint likelihood of consideration and choice given by the dual-goal formulation.

\section{Application 3: Location Choice as a Matrix Factorization Problem}
The formulation for application 3 is a single model but considered in two subsections below. The first subsection describes the matrix factorization formulation. The second subsection gives additional details on the duality between location choice and bid-rent formulations of the home location choice problem.
\subsection{Matrix Factorization Formulation}
Matrix factorization is an approach to decompose the relationship between decision-makers and alternatives via a set of latent vectors. Notation for this model is provided in Table \ref{tab:matrix_factorization}.

\begin{table}[h]
  \caption{Parameters of the matrix factorization model}
  \label{tab:matrix_factorization}
  \centering
\begin{tabular}{cl}
\toprule
$\theta_i$ & Latent preference vector for individual $i$ \\
$\alpha_j$ & Latent attribute vector for alternative $j$ \\
$\epsilon_{ij}$ & Error term for individual $i$ and alternative $j$
assumed to be distributed extreme value type I \\
$U_{ij}$ & Utility associated to alternative $j$ by individual $i$ \\
\bottomrule
\end{tabular}
\end{table}

Consider a utility describing the relationship between individual $i$ and alternative $j$ as $U_{ij}$ and given by
\begin{equation}
    U_{ij} = \theta_i^T \alpha_i + \epsilon_{ij}
\end{equation}
where $\theta_i$ and $\alpha_j$ are latent vectors measuring individual latent preferences and alternative latent attributes, respectively. The model is specified as a hierarchical Bayesian logit model, wherein attributes of the individual and alternative shift the distribution of their respective latent variable vector. The model is estimated by \citet{Athey_Blei_Donnelly_Ruiz_Schmidt_2018} using variational inference to reduce computational costs.

\subsection{Location Choice-Bid Auction Equivalence}
There are two main modeling frameworks  to capture home location choices. First, random utility theory developed from the utility maximization framework, with individuals (households) selecting their preferred location \citep{Lerman_1976}. Second, bid-rent theory in which locations are allocated to individuals (households) based on an auction process \citep{Alonso_1964}. Both frameworks rely on the same utility maximization assumption but differ in their perspective. The location choice approach models individuals as the decision-making agents, selecting among alternative locations. It assumes exogenous prices and that the households are \emph{price takers}. The bid-rent approach models locations as the decision-making agents, selecting among alternative bidding households. However, the two approaches can be shown to represent a \emph{double-matching problem} via the bid-choice equivalence \citep{Martinez_1992}. Notation for this model is provided in Table \ref{tab:bid_choice}.

\begin{table}[h]
  \caption{Parameters and variables of the bid-choice models}
  \label{tab:bid_choice}
  \centering
\begin{tabular}{cl}
\toprule
$z_j$ & Vector of attributes describing alternative location $j$ \\
$U_{ij}$ & Utility associated to alternative $j$ by individual $i$ \\
$W_i(z_j,\bar{U}_i)$ & Willingness-to-pay for location $j$ with reservation utility $\bar{U}_i$ \\
$p_j$ & Price paid for location $j$ \\
$CS_{ij}$ & \makecell[l]{Consumer surplus between the willingness-to-pay by household $j$ and the price paid\\for location $j$} \\
\bottomrule
\end{tabular}
\end{table}

Consider that the location choice approach assumes that households are sensitive to price, which is determined via an auction in the bid-rent approach. Assuming market clearance, the exogenous prices in the location choice approach are equal to the willingness-to-pay (highest bid) in the bid-auction approach. Formally, the location choice problem is defined by
\begin{equation}
\max_{i\in J}{U_{ij}}
\end{equation}
where $U_{ij}$ is the household $i$’s utility function for alternative home location $j$ maximizing utility across $J$ alternatives. In a similar way, the bid-auction problem is defined by

\begin{equation}
i_j^*=\max_{i\in B_j}W_i(z_j,\bar{U}_i)
\end{equation}
where $i_j^*$ is the winning bid in the auction for location $j$, $\bar{U}_i$ is a reservation utility assumed equal across all bids by households $i$, $z_j$ is a vector of attributes for alternative location $j$, and $W_i(z_j,\bar{U}_i)$ is the willingness-to-pay. With these definitions, the bid-choice equivalence can be demonstrated using the following consumer surplus measure
\begin{equation}
CS_{ij}=W_i\left(z_j,{\bar{U}}_i\right)-p_j
\end{equation}
where $CS_{ij}$ is the consumer surplus between the willingness-to-pay and the price $p_j$ paid for a location. It can then be shown that the price $p_j$ is defined by the maximum willingness-to-pay among all bidding households, giving the following identity

\begin{equation}
j_i^* = \argmax_{j\in J} (W_i(z_j, \bar{U}_i) - \max_{g \in B_j\subset C} w_g(z_j,\bar{U}_g)) 
\end{equation}
where $B_j\subset C$ is the set of bidders, the inner process represents the bid-auction process capturing the maximum bid for location $g$, and the outer process represents the location choice for household $i$. The relationship will be equal to zero if the household locates at the location and negative if they are outbid (i.e., the maximum bid is greater than their willingness-to-pay).

\end{document}